\documentclass[runningheads,a4paper]{llncs}

\usepackage{amssymb}
\usepackage{graphicx}
\usepackage{xcolor}

\usepackage{xurl}
\usepackage{hyperref}
\newcommand{\keywords}[1]{\par\addvspace\baselineskip
\noindent\keywordname\enspace\ignorespaces#1}

\begin{document}

\mainmatter  % start of an individual contribution

\title{Towards a Simple and Extensible Standard for Object-Centric Event Data (OCED) --- Core Model, Design Space, and Lessons Learned}

\titlerunning{Towards Simple and Extensible OCED}

\author{
Dirk Fahland\inst{1}\and
Marco Montali\inst{2}\and
Julian Lebherz\inst{3}\and
Wil M.P. van der Aalst\inst{4}\and
Maarten van Asseldonk\inst{5}\and
Peter Blank\inst{6}\and
Lien Bosmans\inst{7}\and
Marcus Brenscheidt\inst{8}\and
Claudio di Ciccio\inst{9}\and
Andrea Delgado\inst{10}\and
Daniel Calegari\inst{10}\and
Jari Peeperkorn\inst{11}\and
Eric Verbeek\inst{1}\and
Lotte Vugs\inst{5}\and
Moe Thandar Wynn\inst{12}\thanks{Moe Thandar Wynn and Julian Lebherz coordinated the efforts of the OCED working group 2021-2024 which led to the results presented in this report.}
}
\authorrunning{Fahland, Montali, Lebherz et al.}

% affiliations
\institute{
%1
Eindhoven University of Technology, the Netherlands, \email{\{d.fahland,h.m.w.verbeek\}@tue.nl}\and
%2
Free University of Bozen-Bolzano, Italy, \email{montali@inf.unibz.it}\and
%3
Standard Chartered Bank, Singapore, \email{julian.lebherz@sc.com}\and
%4
Process and Data Science Chair, RWTH Aachen University, Aachen, Germany, \email{wvdaalst@pads.rwth-aachen.de}\and
%5
Konekti, Eindhoven, the Netherlands, \email{\{maarten,lotte\}@getkonekti.io}\and
%6
PwC, Switzerland, \email{peter.blank@pwc.ch}\and
%7
Randstad Digital, Leuven, Belgium, \email{lienbosmans@live.com}\and
%8
\email{Mbrenscheidt@outlook.de}\and
%9
Utrecht University, The Netherlands, \email{c.diciccio@uu.nl}\and
%10
Instituto de Computación, Facultad de Ingeniería, Universidad de la República, Uruguay, \email{\{adelgado,dcalegar\}@fing.edu.uy}\and
%11
Research Center for Information Systems Engineering (LIRIS), KU Leuven, Belgium, \email{jari.peeperkorn@kuleuven.be}\and
%12
Queensland University of Technology, Australia, \email{m.wynn@qut.edu.au}
}

%\toctitle{Lecture Notes in Computer Science}
%\tocauthor{Authors' Instructions}
\maketitle

\begin{abstract}
Process mining is shifting towards use cases that explicitly leverage the relations between data objects and events under the term of \emph{object-centric process mining}. Realizing this shift and generally simplifying the exchange and transformation of data between source systems and process mining solutions requires a standardized data format for such \emph{object-centric event data} (OCED). This report summarizes the activities and results for identifying requirements and challenges for a community-supported standard for OCED. (1) We present a proposal for a \emph{core model} for object-centric event data that underlies all known use cases. (2) We detail the limitations of the core model wrt. a broad range of use cases and discuss how to overcome them through conventions, usage patterns, and extensions of OCED, exhausting the \emph{design-space for an OCED data model} and the inherent trade-offs in representing object-centric event data. (3) These insights are backed by \emph{five independent OCED implementations} which are presented alongside a series of lessons learned in academic and industrial case studies. The results of this report provide guidance to the community to start adopting and building new process mining use cases and solutions around the reliable concepts for object-centric event data, and to engage in a structured process for standardizing OCED based on the known OCED design space.

\end{abstract}

%%% =============================================================
\section{Introduction}
\label{sec:rationale_oced}
%%% =============================================================

\subsubsection{Rationale behind OCED}
In the past decade, Process Mining has not only seen tremendous growth in the academic arena, but also started to establish itself as one of the predominant approaches to improving processes for larger companies of virtually any industry. Process Mining and related services have become a sizeable business - for software vendors, professional service firms, and commercial end user alike.

Such a trajectory naturally spurs substantial investment and advancement; however, it is also typical that\,--\,in an attempt to safeguard intellectual property\,--\,many new features, products and services are being shielded off from usage by other players in the ecosystem. Especially when this affects areas that are of concern to all market participants, such silos impede competition, innovation and the pace of further development. In many industries that are heavily reliant on the exchange of something, standardizing terms and conditions of such exchange led to a great leap forward for the entire ecosystem – think containers for global trade, internet protocol for global communication.

Process Mining itself is heavily reliant on the exchange of data, which typically originates from systems that were not designed around this use case and hence requires substantial transformation. Market participants have created different approaches to reduce the effort required for data transformation, but so far no data exchange format has seen enough adoption to be nominated as a de-facto standard. The relevance and magnitude of this bottleneck as one of the predominant effort drivers in process mining projects has been reconfirmed by the \emph{IEEE Task Force on Process Mining} (TFPM)\footnote{\url{https://www.tf-pm.org/}}.

With the \emph{IEEE eXtensible Event Stream (XES)}~\cite{DBLP:conf/caise/VerbeekBDA10_xes,DBLP:journals/cim/AcamporaVSAGV17_xes,DBLP:journals/cim/WynnAVS24_xes} initiated in 2010, academia has established a data exchange format, which fueled tremendous growth in process mining research. 
Standardizing how tools capture, transfer, load, and interpret event data continues to pay dividends. 
XES is supported by several commercial tools, and all tools support the main concepts identified in the XES meta-model (e.g., concepts like event, trace, timestamp, and attribute).
However, the specific XML file format is not widely used in practice.
Moreover, process mining adoption by business entities and scientific progress in the past decade changed the requirements towards an up-to-date standard substantially. 
The discipline is moving from case-centric event data to object-centric event data~\cite{DBLP:conf/sefm/Aalst19}, making XES less relevant.
Therefore, the IEEE TFPM decided in 2021 to initiate a community process to co-design the XES successor \textbf{Object-Centric Event Data (OCED)} as a data exchange format to standardize and thereby facilitate system interoperability within process mining and adjacent areas, like process automation, process simulation, and Business Process Management. OCED shall:
\begin{enumerate}
    \item \textbf{spur innovation and competition} by lowering the barrier for new ideas and market participants to access and enter the eco-system
    \item \textbf{improve the Return on Investment (ROI)}
    \begin{enumerate}
        \item \textbf{for all market participants} by improving the security of investments with standardized eco-system access,
        \item \textbf{for commercial end users and professional service firms} by allowing them to focus more on business value creation and less on data transformation,
        \item \textbf{for software vendors} by allowing them to focus on differentiating functionality rather than developing yet another data transformation engine, and
    \end{enumerate}
    \item \textbf{create a new marketplace} for source system-specific adapter modules, translating source data into/from OCED.
\end{enumerate}

\subsubsection{Balancing Simplicity and Expressivity}
This report summarizes the community efforts and results in developing proposals for establishing a new standard for Object-Centric Event Data. Section~\ref{sec:path_to_oced} briefly documents the events and activities organized by the IEEE Task Force on Process Mining in eliciting requirements and proposals for OCED. A key challenge emerging from this process was that OCED has to be, both:
\begin{enumerate}
    \item conceptually simple to facilitate implementation and adoption in practice, and
    \item conceptually expressive to allow supporting a large number of use cases (considering both forms and types of data in source systems and analysis objectives over this data)
\end{enumerate}
Feedback on initial proposals for OCED (included in Appendices~\ref{app:oced-mm-original} and \ref{app:ocel2}) shared with the community led to the consensus that any standard model for OCED should start from a \emph{core} of essential concepts necessary in all use cases. The core in turn should be easily extensible to cover a wider range of use cases.

Section~\ref{sec:oced_core} documents the proposal for such a core model for OCED, that can be seen as the common denominator of the initial proposals for OCED, as well as its limitations. 

Section~\ref{sec:challenges} summarizes the arguments made around the need for extending OCED beyond its core as well as the implications and challenges when one attempts to do so. These challenges concern basic principles of conceptual modeling and data modeling as well as implementation considerations. Overcoming these challenges requires to establish further conventions and best practices around OCED. 

Section~\ref{sec:implementations} complements these challenges with reports on concrete results and experiences gathered from four independent implementations of OCED (its core and extensions). Besides highlighting several practical challenges in implementing OCED and how these have been overcome, this section also provides pointers to resources for working with OCED and its further development.

The concluding Section~\ref{sec:conclusion} summarizes how the core model for OCED forms a reliable basis for building an eco-system of process mining around object-centric event data but also the potential sources of ambiguity that still hinder inter-operability. It also outlines the open challenges and a possible roadmap to make OCED more robust for advanced use cases.

This report thus summarizes the lessons learned on identifying core concepts of OCED and its extensions, but does not formulate a standard\footnote{This document deliberately does not refer to any of the data models or meta-models for OCED presented in this report as ``standard'' as these currently do not meet the associated requirements.} for OCED. Rather, its contents shall inform the next steps in the process of developing a community standard for OCED.

%%% =============================================================
\section{The Path to OCED}
\label{sec:path_to_oced}
%%% =============================================================

The community process initiated by the IEEE TFPM on co-designing OCED involved several steps.

\subsubsection{Requirements gathering.}

The requirements for OCED were gathered through an online survey with 289 participants and a XES 2.0 workshop co-located with the 3rd International Conference on Process Mining (Eindhoven, 2021)~\cite{DBLP:conf/icpm/WynnLAACJV21_xes_survey}. This led to the following three observations
\begin{enumerate}
\item the single-case-notion is very limiting, leading to a disconnect between reality and the represented events,
\item the XES standard is too complex and many of its extensions are rarely used, and
\item XES is associated with a particular XML storage format making it impractical for many real-life use cases. 
\end{enumerate}
Learning from the problems associated with XES, OCED needs to be 
\begin{enumerate}
    \item object-centric (i.e., an event may refer to any number of objects instead of a single case) \cite{DBLP:conf/sefm/Aalst19,Aalst_2023_ocpm_unraveling,DBLP:reference/bdt/Fahland19}, 
    \item as simple as possible, and
    \item have a meta-model decoupled from a particular storage format. 
\end{enumerate}

\subsubsection{Proposal development.}

A core \emph{OCED working group} of eight experts from academia and industry was formed to develop an initial proposal for the \emph{OCED Meta-Model} (OCED-MM). Diverse views and opinions were solicited and discussed among the core team in striking the right balance between expressivity of the model and its simplicity. The resulting meta-model captures the concepts that have a majority vote. The meta-model resulting from this discussion, consisting of a \emph{Base Model} and a \emph{Full Model}, is included in Appendix~\ref{app:oced-mm-original}. 

It was circulated for feedback in two runs: a first run on August, 2nd 2022 to respondents of the 2021 survey, registered attendees of the 2021 XES Workshop as well as the TFPM steering committee and advisory board, and a second run on September, 9th 2022 to all subscribers of the TFPM newsletter. 

As part of the 4th International Conference on Process Mining (Bolzano, 2022) an XES Symposium  was held, in which the OCED-MM was presented and discussed. Many interesting thoughts and ideas had been brought to the attention of the OCED working group during this meeting; specifically, symposium participants stated that OCED-MM was, both,
\begin{itemize}
\item not expressive enough as it does not natively support a number of use cases considered relevant by the participants, essentially asking to extend OCED-MM further, and
\item being too complex in the concepts that need to be considered and implemented, hindering adoption.
\end{itemize}
This feedback underlined the difficulty of striking the right balance between expressivity and simplicity. Yet, the core concepts of the OCED-MM Base Model remained in consensus.

\subsubsection{Call for Reference Implementations}

In order to get further clarity and feedback on the effort and challenges in implementing an object-centric event data model and exchange format the OCED working group issued on March 10, 2023 a \emph{Call for Reference Implementations} of the OCED-MM Based Model or the OCED-MM Full Model.

As part of the 5th International Conference on Process Mining (Rome, 2023), an OCED Symposium was held at which 4 independent implementations were presented and discussed:
\begin{itemize}
\item an implementation of the \emph{OCED-MM Base Model} by the company Konekti\footnote{\url{https://getkonekti.io/}}, further described in Sect.~\ref{sec:implementation:konekti};
\item an implementation of the \emph{OCED-MM Base Model} called OpenOCED by Delgado et al.~\cite{DBLP:conf/bpm/Calegari023}, further described in Sect.~\ref{sec:implementation:open_oced};
\item an implementation of the \emph{OCED-MM Full Model} by Swevels et al.~\cite{DBLP:conf/icpm/SwevelsFM23_oced-pg}, further described in Sect.~\ref{sec:implementation:oced-pg}; and
\item an implementation of a variation of the OCED-MM Full Model called \emph{Object Centric Event Log V2 (OCEL 2.0)}\footnote{\url{https://www.ocel-standard.org/}} by Koren et al.\ \cite{ocel2_specification,DBLP:conf/icpm/KorenABA23}; the core ideas of the OCEL 2.0 model are shown in App.~\ref{app:ocel2}, the implementation is further described in Sect.~\ref{sec:implementation:ocel2}.
\end{itemize}
A fifth independent implementation has recently been published:
\begin{itemize}
\item an implementation of the \emph{OCED-MM Full Model} and \emph{OCEL 2.0} by Bosmans et al. called \emph{Stack't}, further described in Sect.~\ref{sec:implementation:stack-t}
\end{itemize}
Alongside the implementations, several datasets in OCED, OCEL 1.0, and OCEL 2.0 format were made available to the community for exploration and adoption (see Sect.~\ref{sec:implementations}).

While the independent implementations confirmed that the core ideas of OCED-MM are viable, the discussion among the presenters and participants of the OCED symposium noted a need for, both, 
\begin{itemize}
\item clarifying the relations and compatibility between the different implementations, and
\item clarifying the core concepts of the OCED-MM Base Model and the extensions in the OCED-MM Full Model and OCEL 2.0,
\end{itemize}
before the proposal can enter a formal standardization process. 

In order to facilitate this clarification and pave the way towards standardization, the OCED working group concluded to:
\begin{enumerate}
\item derive a minimal \emph{OCED-MM Core Model} which lies at the intersection of all prior  proposals and implementations while supporting all essential core concepts, and
\item clarify towards the community the lessons learned and open challenges in extending this OCED-MM Core Model to more expressive OCED models and their implementation.
\end{enumerate}
The remainder of this document presents the results of these efforts and discussions. Section~\ref{sec:oced_core} describes the OCED-MM Core Model. Section~\ref{sec:challenges} details the challenges in implementing and extending the OCED-MM Core Model. Section~\ref{sec:implementations} summarizes the lessons learned of implementing OCED in the independent implementations.

%%% =============================================================
\section{OCED Meta-Model - Core Model}
\label{sec:oced_core}
%%% =============================================================

\begin{figure}
    \centering
    \includegraphics[width=\linewidth]{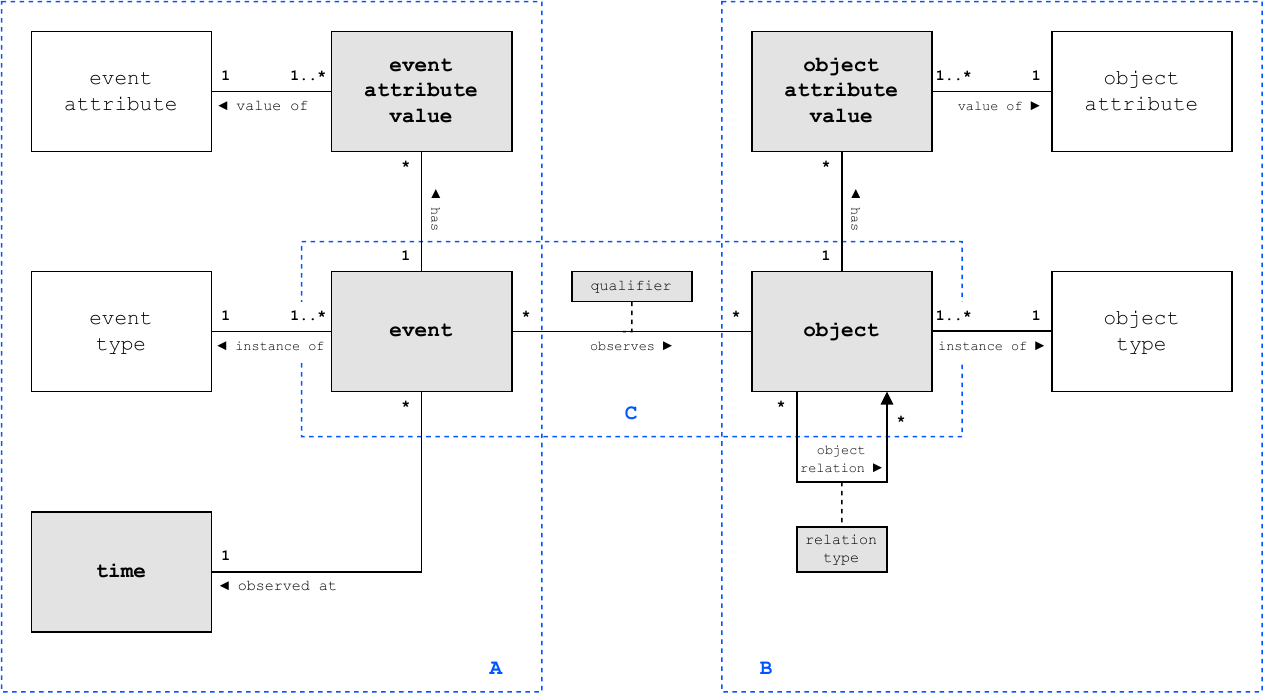}
    \caption{OCED-MM Core Model}
    \label{fig:oced-mm:core}
\end{figure}

The OCED-MM Core Model shown in Fig.~\ref{fig:oced-mm:core} is the simplest\footnote{in terms of number of concepts and relations included in the model} model for object-centric event data that the OCED working group could identify. Section~\ref{sec:oced_core:concepts} describes the model which is illustrated by an example in Section~\ref{sec:oced_core:example}. Section~\ref{sec:oced_core:interpretation} highlights non-trivial aspects in the interpretation of the \emph{object} concept in OCED. 

Section~\ref{sec:oced_core:limitations} lists the known limitations of this minimal model, also highlighting the interests of the process mining community for more expressivity in an OCED model.

%%% -------------------------------------------------------------
\subsection{Core Model Concepts}
\label{sec:oced_core:concepts}
%%% -------------------------------------------------------------

In order to present the OCED-MM Core Model concepts in manageable chunks, Figure~\ref{fig:oced-mm:core} is structured into three groups highlighted in blue. 
\begin{itemize}
    \item \textcolor{blue}{Group A} describes events, associated attributes and the time construct, which can be closely related to the XES standard~\cite{DBLP:journals/cim/WynnAVS24_xes} and many non-object centric process mining solutions in the market.
    \item \textcolor{blue}{Group B} describes the objects, associated attributes, as well as object relations. Inclusion of these concepts directly stems from real-world requirements collected during the development of the proposal, see Sect.~\ref{sec:path_to_oced}.
    \item \textcolor{blue}{Group C} connects the concepts of events and objects in groups A and B.
\end{itemize}

\subsubsection{Events.} Starting with \textcolor{blue}{Group A}, the following event-centric concepts are captured.

\subsubsection{1. event} \label{subsubsec:event}
| An event describes the occurrence of an observable phenomenon. An event is atomic, meaning it refers to an observation taking place at exactly one point in time rather than having a duration. Every event has, viz. is \texttt{instance of}, exactly one \texttt{event type} (e.g., \emph{[purchase order approved]}; captured as string). It denotes the kind of observation described by the event. In most use cases, the event type is the process \emph{activity} that was performed, though other types of observations can be described as well (e.g., sensor recordings). Each defined \texttt{event type} has at least one event that instantiates it (e.g., \emph{[purchase order approved]} was observed in at least one event \emph{[purchase order ID\#298374 approved]}). 

\subsubsection{2. time} |  \texttt{Time} describes the moment in time where the event has been \texttt{observed at}. It captures both a timestamp conforming to ISO 8601-1:2019 and its resolution (ref. to the precision in which the timestamp was recorded). At a minimum, the following precisions are to be differentiated: date, hour, minute, second, millisecond. If the time zone is omitted, all timestamps are treated as UTC. While the meta-model does not prescribe a method to represent the timestamp-resolution-pair, the ISO standard 8601-2:2019 proposes uncertainty classifiers as a way to store both components in one string.

\subsubsection{3. event attribute value} |  Each event \texttt{has} an arbitrary number of \texttt{event attribute values} (e.g., \emph{[USD]} and/or \emph{[Emirates Airlines]}) and corresponding \texttt{event attribute} names (e.g., \emph{[transaction currency]} and/or \emph{[merchant name]}) further describing the observation captured by the event as \emph{attribute-value pairs} (e.g., \emph{[transaction currency] = [USD]} and/or \emph{[merchant name] = [Emirates Airlines]}). Each \texttt{event attribute value} is captured as string, boolean, integer, real, date, time or timestamp (the latter three in accordance with ISO 8601-1:2019). 
Each event attribute value is related to exactly one event (i.e., if two different events have attributes with the same value, then each event has its ``own copy'' of the value). Each defined event attribute name relates to at least one event attribute value  (e.g., \emph{[user type] = [manual]} or \emph{[RPA]} or \emph{[automated]}), and every event attribute value is \texttt{value of} exactly one event attribute name (i.e., if two different attributes have the same value, then each attribute name has its ``own copy'' of the value). Some information is typically represented with value-unit-pairs (e.g., price and currency) describing parts of the same logically connected information. In such cases, it is good practice to indicate their relation by choosing the unit’s event attribute name as the value’s event attribute name suffixed with “\_unit” (e.g., \emph{[price] = [48.76]} and \emph{[price\_unit] = [USD]}). Please note that this convention is not prescribed by the meta-model. 

\subsubsection{Objects and Relations.} 
In \textcolor{blue}{Group B}, the following object-centric concepts are captured:

\subsubsection{4. object} \label{subsubsection:object}
|  An \texttt{object} either represents something tangible or abstract. Examples of tangible objects are persons, locations, machines, documents, document line items. Examples of abstract objects are legal entities, organizational constructs, and electronic documents. To represent objects in accordance with the model, it is not required to classify them into either of the two. Every object has exactly one object type (e.g., \emph{[sales order]}; captured as string), while each defined object type relates to at least one object (e.g., \emph{[sales order ID\#12345]}).

\subsubsection{5. object attribute value} |  Each object \texttt{has} an arbitrary number of \texttt{object attribute values} (e.g., \emph{[blue]} and/or \emph{[1897]}) and corresponding \texttt{object attribute} names (e.g., \emph{[color variant]} and/or  \emph{[weight]}) further describing the object as \emph{attribute-value pairs} (e.g., \emph{[color variant] = [blue]} and/or \emph{[weight] = [1897]}). Each \texttt{object attribute value} is captured as string, boolean, integer, real, date, time or timestamp (the latter three in accordance with ISO 8601-1:2019). Nested object attribute values are not supported. Each \texttt{object attribute value} is related to exactly one object (i.e., if two objects have the same attribute value, then each object has its ``own copy'' of this value). Every object attribute value is \texttt{value of} exactly one object attribute name, while each defined object attribute name relates to at least one object attribute value (i.e., if two different attributes have the same value, then each attribute name has its ``own copy'' of the value). Some information is typically represented with value-unit-pairs describing parts of the same logical information. In such cases, it is good practice to indicate their relation by choosing the unit’s object attribute name as the value’s object attribute name suffixed with “\_unit” (e.g., \emph{[weight] = [1897]} and \emph{[weight\_unit] = [kg]}). Please note that this convention is not prescribed by the meta-model.

\subsubsection{6. object relation} \label{subsubsection:object_relation}
| An \texttt{object relation} represents a link between a pair of \texttt{objects} and is represented as as \emph{directed relationship} from one \texttt{object} (reflecting the relation's origin) pointing to one \texttt{object} (reflecting the relation's target). Every object relation has exactly one object \texttt{relation type} (e.g., \emph{purchase order line items} being related to their parent \emph{purchase order}; their object relation type is \emph{[child of]}; captured as string). Possible relation types are \texttt{CHILD\_OF} and \texttt{PARENT\_OF}) but these are not prescribed by the core model and additional object relation types can be introduced as part of the data capture and used to reflect semantics of the relation.

\subsubsection{Event-to-Object Relations.} In \textcolor{blue}{Group C}, events and objects are connected.

\subsubsection{7. observes relation from events to objects} \label{subsubsection:observes_relation_from_events_to_objects}
| In an object-centric setting, an \texttt{event} not only observes an activity, viz. \texttt{event type}, but also explicitly \texttt{observes} (changes to) \texttt{objects}. An \texttt{even}t and an \texttt{object} can be related in a \texttt{qualified} (i.e., association class) manner, meaning their type of relationship is denoted. Possible qualifiers are \texttt{CREATE}, \texttt{MODIFY} and \texttt{DELETE} but these are not prescribed by the core model and additional qualifiers can be introduced as part of the data capture and used to reflect the semantics of the relationship. Each object can be observed by an arbitrary number of events, while each event can observe an arbitrary number of objects. This means there can be events without objects and vice-versa.

%%% -------------------------------------------------------------
\subsection{Example}
\label{sec:oced_core:example}
%%% -------------------------------------------------------------

The following shall exemplify how the core meta-model can be applied. For this, real-world scenarios in the context of a purchase-to-pay process have been selected. 

\subsubsection*{A) PO creation} 
\begin{figure}
    \centering
    \includegraphics[width=\linewidth]{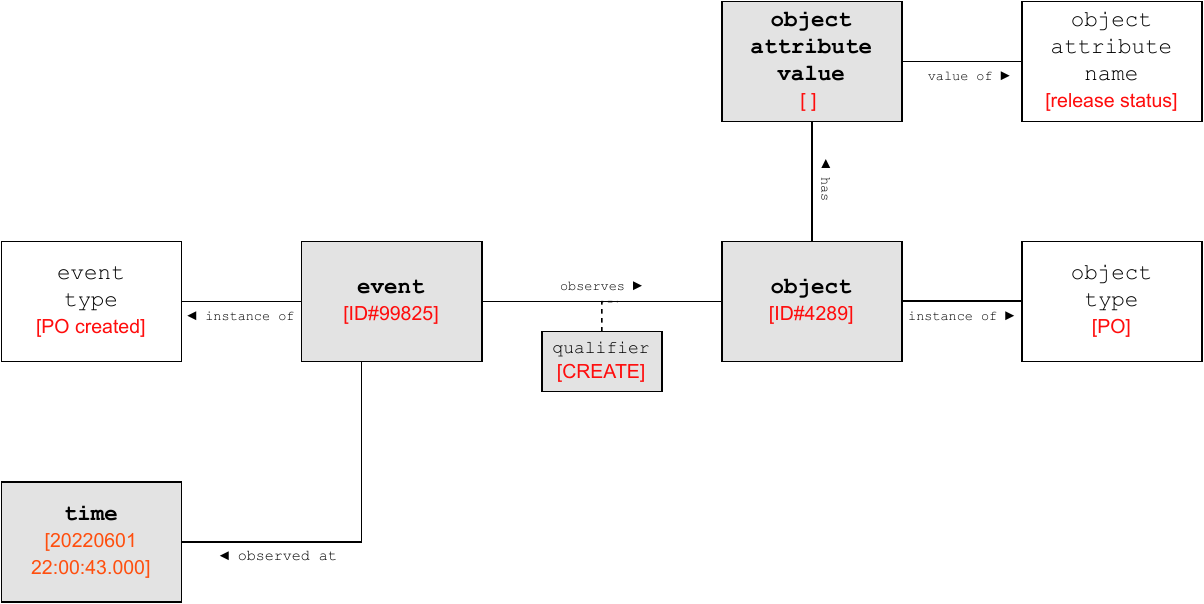}
    \caption{Example Scenario A - PO created}
    \label{fig:example:po_created}
\end{figure}

| A purchase order document is created. In this context we focus on the PO header information only. Hence a PO object is created alongside a number of object attributes. For simplicity we solely present the PO’s release status, which is created as non-released. The respective values are marked with red font in Figure~\ref{fig:example:po_created}, sparing unused parts of the meta-model.

\subsubsection*{B) PO release} 
\begin{figure}
    \centering
    \includegraphics[width=\linewidth]{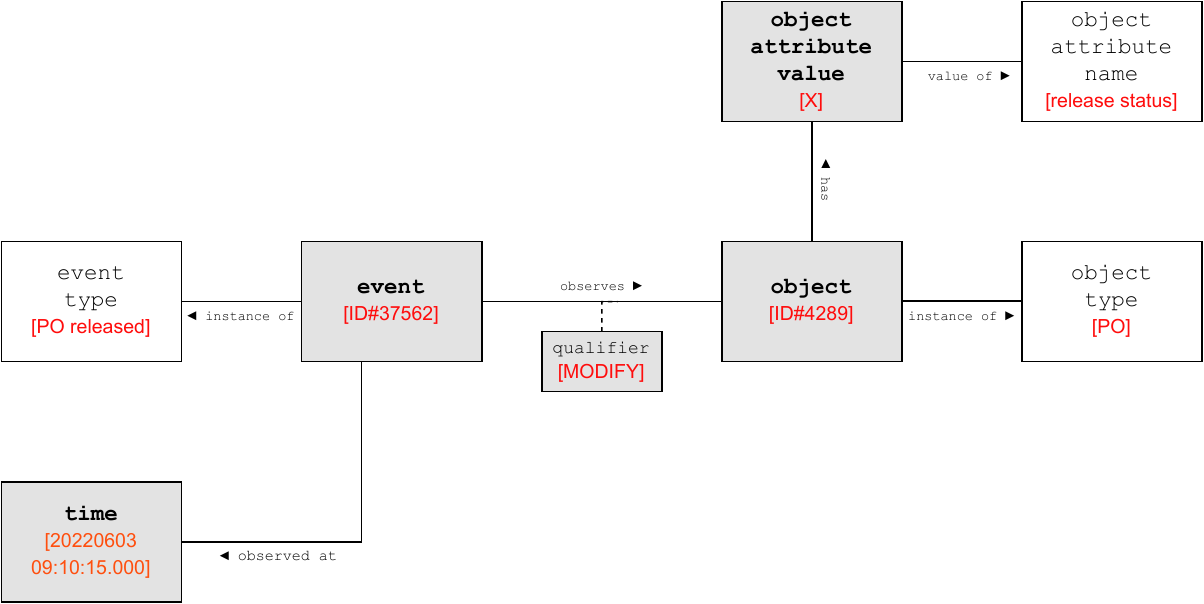}
    \caption{Example Scenario B - PO released}
    \label{fig:example:po_released}
\end{figure}

| The purchase order is released. Data-wise this is reflected in an updated release status. The object itself does not change, but its object attribute value does. Please refer to Figure~\ref{fig:example:po_released} for details.

\subsubsection*{C) Invoice Receipt} 
\begin{figure}
    \centering
    \includegraphics[width=\linewidth]{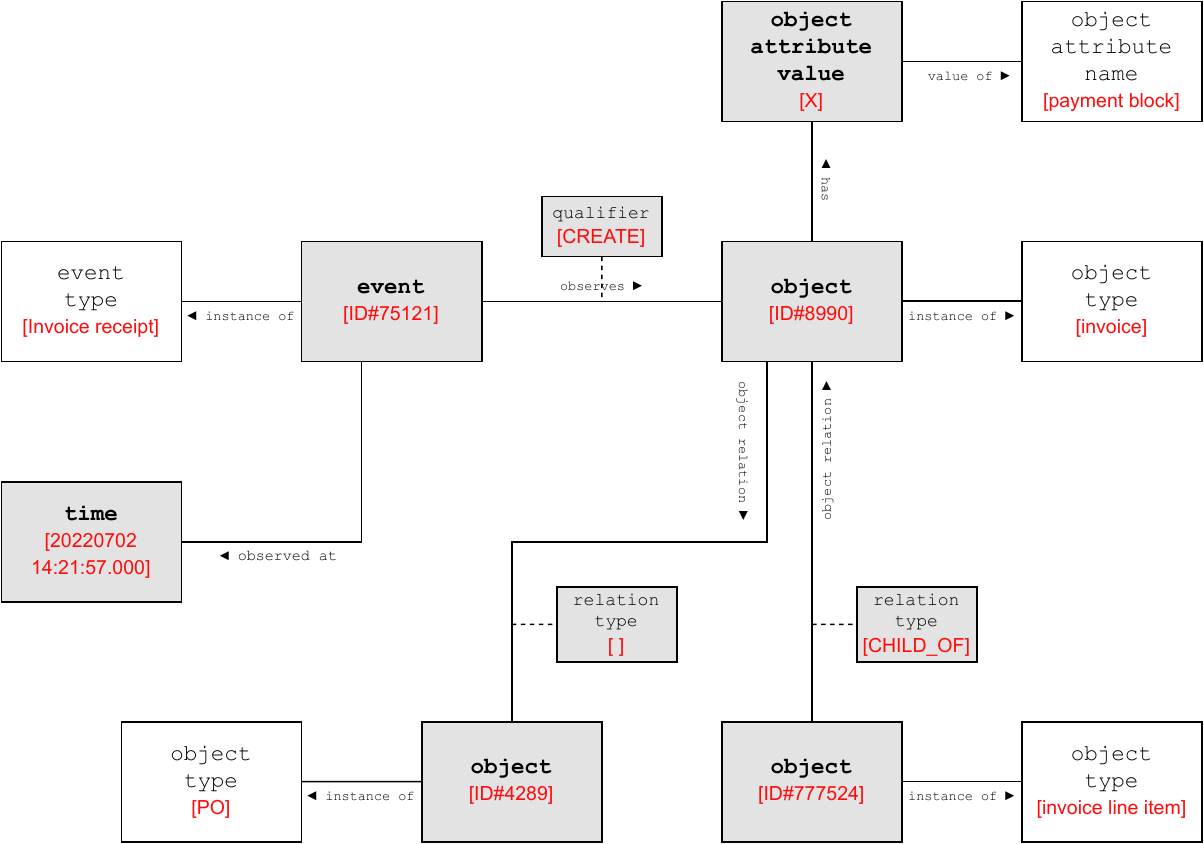}
    \caption{Example Scenario C - Invoice receipt}
    \label{fig:example:invoice_receipt}
\end{figure}

| An invoice is received in relation to the PO created in scenario A. The invoice, one of its invoice line items, as well as their object relation get recorded. Another object relation is recorded for the link to the PO. Further PO details (i.e., object attribute values) of the purchase order are omitted in Figure~\ref{fig:example:invoice_receipt} since they do not change. As per standard practice this invoice gets recorded with an active payment block. Since this object attribute value is associated to the invoice (header), it implicitly applies to its children, i.e., invoice line items, as well.

%%% -------------------------------------------------------------
\subsection{Baseline Interpretation of OCED}
\label{sec:oced_core:interpretation}
%%% -------------------------------------------------------------

In OCED, events and objects have distinct roles: an event describes an observation in time\footnote{While most events describe observations recorded in the past, an event may  also describe a ``future observation'' such as a due date of an invoice}, whereas an object describes a tangible or abstract entity that is observed (i.e., via the qualified \texttt{observes} relationship). 
\begin{itemize}
    \item Each object needs a \emph{unique object identifier} that events and relations can refer to.
    \item Likewise, each event needs a \emph{unique event identifier} that objects can refer to, but only events carry a timestamp.
    \item Attributes are ``owned'' by their parent concept (object, event) and thus should always be represented/serialized as children of their parent concept that do not carry their own identifier.
    \item Relations in the OCED-MM Core Model are not distinct entities carrying their own identifier; see Sect.~\ref{sec:oced_core:limitations}.
\end{itemize}

\begin{figure}
    \centering
    \includegraphics[width=\linewidth]{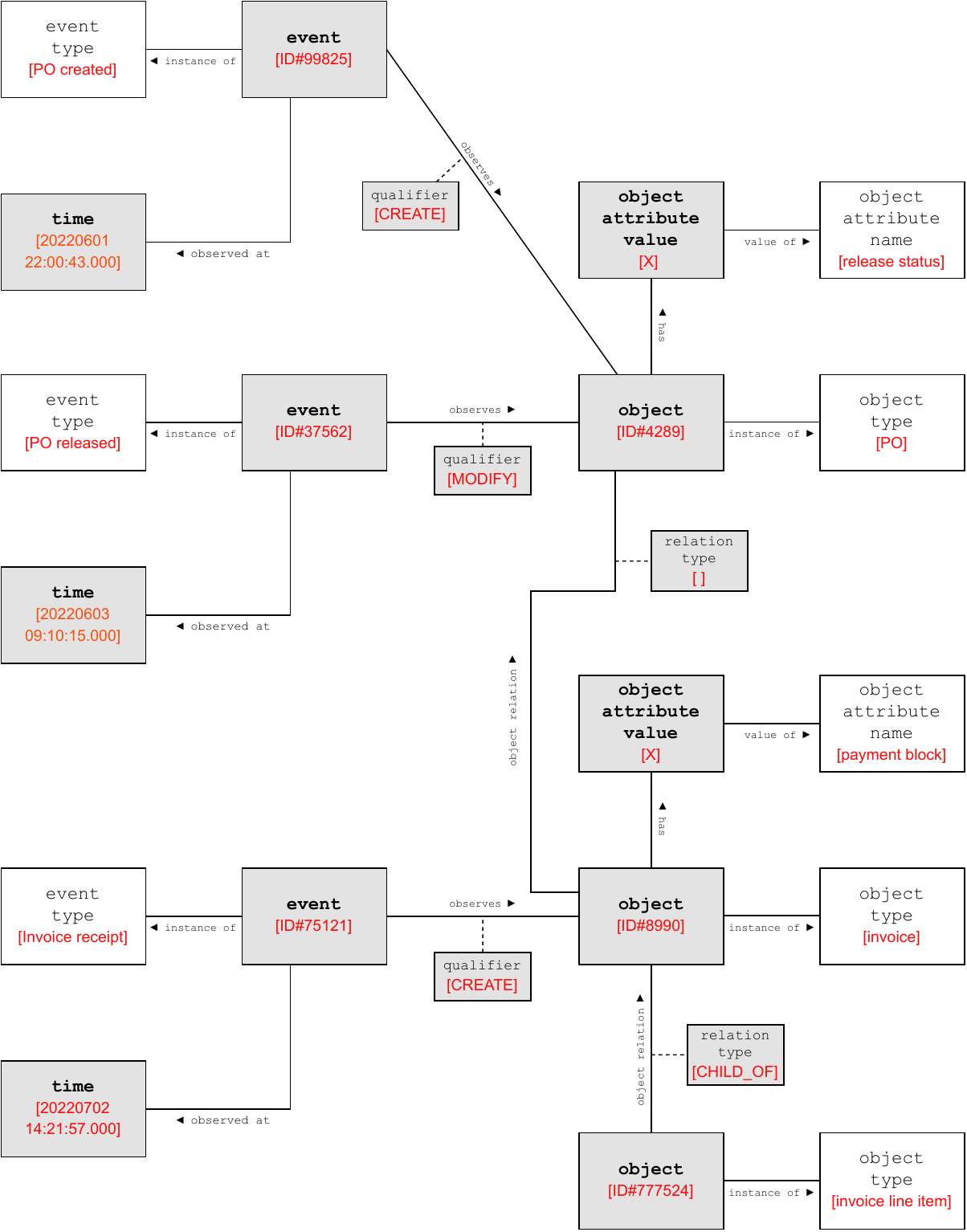}
    \caption{Baseline representation and interpretation of the Scenarios A-C in a single OCED instance}
    \label{fig:example:minimal}
\end{figure}

The current OCED-MM Core Model allows some room for interpreting these concepts. The following is a ``minimal'' \emph{baseline interpretation} of the concepts that can be understood as the basis for OCED and all further extensions.
\begin{itemize}
    \item In an OCED instance (i.e., a concrete dataset), each object, relation, and attribute is represented once, describing a single state or static view on all objects, relations, and attributes at an \emph{unspecified} point in time. This could be the state of all objects, relations, and attributes at the time of extraction of the data, but this is subject to how the source system provides the data and the type of extraction. Therefore, the OCED Core Model does not enforce this interpretation.
    \item An \texttt{observes} relation from an event $e$ to an object $o$ is essentially only a reference of $e$ to the identifier of $o$ denoting that $e$ observed or operated on $o$ but not denoting what of $o$ has been observed or changed or in which state $o$ has been.
    \item In this representation, the semantics of the event $e$ wrt. $o$, i.e., what $e$ did with $o$, lies in the qualifier of the \texttt{observes} relation, e.g., CREATE or a domain-specific qualifier, and in domain-knowledge about $e$, e.g., the event‘s type being \emph{[Create Order]} or \emph{[Change Price]} or \emph{[Re-assign flight]}.
\end{itemize}
Figure~\ref{fig:example:minimal} visualizes this baseline interpretation on an OCED instance that combines all three scenarios A-C from Sect.~\ref{sec:oced_core:example}. Note that this representation shows the latest \texttt{value} of \texttt{attribute} \emph{release status} of the PO \texttt{object} (after \emph{PO released}), but not its initial value (before \emph{PO released} occured).
While this baseline interpretation limits which process dynamics OCED can describe without further domain knowledge, it arguably contains the minimum requirements for all interpretations of OCED: 
\begin{itemize}
\item each object, relation, attribute is described, 
\item each event is described, 
\item the existence of relations between events and objects is stated
\end{itemize}
in the minimum required form allowing usage and extension in a variety of use cases. 

Note that the OCED-MM Core Model does not prescribe this baseline interpretation, e.g., it does not prescribe that an object is only represented once in the data or that event semantics must be deferred to domain knowledge. However, richer representations and interpretations of data\,--\,such as also describing the different values of the \emph{release status} attribute\,--\,require more care and consensus as discussed next.

%%% -------------------------------------------------------------
\subsection{Known Limitations}
\label{sec:oced_core:limitations}
%%% -------------------------------------------------------------

The meta-model shall be kept as simple as possible, while retaining relevance in industry and academia. This implies that there are some known limitations (and corresponding workarounds) that are also further discussed in Sect.~\ref{sec:challenges}: 

\begin{enumerate}
    \item \textbf{Event atomicity \& no relations between events} | An event is atomic (i.e has exactly one timestamp) and it is not possible to directly relate events to each other, hence, amongst others, causality relations, activities with both a start and end event and partial orders cannot be stored explicitly; see Sect.~\ref{sec:challenges:other_as_objects}.
    \item \textbf{No complex attribute values} | The data type of any event attribute value and object attribute value are consistent throughout the whole model, i.e., all values of the same attribute name can only be of type string, boolean, integer, real, date, time or timestamp (date, time and timestamp in accordance with ISO 8601-1:2019). Complex (or nested) data types are not supported. This affects representing complex values as well as how units of numeric values are stored; see Sect.~\ref{sec:challenges:attributes_as_objects}.
    \item \textbf{Limited semantics} | Events, object types, and object and event attribute values may carry semantics specific to selected domains. It is encouraged to establish such conventions, however these are not enforced by the meta-model. 
    \item \textbf{Binary object relations} | Object relations in the meta-model are directed and binary. Even though some of the relationships in real life are tertiary or of higher order, such relationships are less frequent; see Sect.~\ref{sec:challenges:relations_as_objects}.
    \item \textbf{Ambiguous representation of object relations} | The meta-model does not prohibit object attributes to store references to other objects, allowing ambiguous and inconsistent representations of object relations, see Sect.~\ref{sec:challenges:relations}.
    \item \textbf{Relations are identified by their objects} | Object relations are defined by their source and target object and their type. While this is sufficient for expressing static data models, i.e., where object relations do not change, changes to object relations (creation, modification, deletion) are not unambiguously expressible, see Sect.~\ref{sec:challenges:relations_as_objects}.
    \item \textbf{Limited types and no data schema} | Each concept (event, object, object relation, attribute) has a \emph{type}, but these types are not related to each other and some domains may ascribe more than one type to them, see Sect.~\ref{sec:challenges:strict_extensions}.
    \item \textbf{Meta-model vs. reference implementations} | The meta-model does not prescribe how things are stored or syntactically represented. For example, there may be one table per event type and one table per object type. The qualified relations may also correspond to tables. Things will be typed, but this is outside the scope of the meta-model and needs to be detailed for reference implementations of this meta-model. 
\end{enumerate}

%%% =============================================================
\section{Challenges in Standardizing and Extending OCED}
\label{sec:challenges}
%%% =============================================================

\subsubsection{Consistent interpretation between producers and consumers.}
The OCED-MM Core Model presented in Sect.~\ref{sec:oced_core} outlines the core concepts for representing object-centric event data for exchange and storage. Adoption of OCED in practice is subject to it fulfilling a role as \emph{intermediary} between \emph{data producers} (e.g., source systems, ETL solutions) and \emph{data consumers} (e.g., process mining algorithms, tools, and solutions) for a variety of use cases. Thereby, a standardized OCED format has to codify to both producers and consumers how various aspects of object-centric event data are to be \emph{represented} and how various representations of object-centric event data are to be \emph{interpreted}\,--\,ultimately allowing a producer to provide a serialization of OCED that a consumer can unambiguously interpret.

\subsubsection{Transport vs. storage vs. analysis.}
The large variety of forms in which object, relations, and event data are stored in source systems, as well as the broad range of analysis use cases pose a series of challenges for standardizing representation and interpretation of OCED. Further, process mining tools consuming OCED must implement OCED-compliant data structures that internally model object-centric event data suitable for querying and algorithms. OCED for data exchange may prioritize fewer concepts to reduce storage footprints (e.g., relying on implicit semantics and conventions). Instead, OCED for storage, querying, or algorithms may prefer more explicit representations using more concepts to improve performance and functionality.

The following sections detail these challenges and also outline how extending the core model introduces further sources of ambiguities that can be resolved through agreeing on conventions in representing and interpreting OCED.

%%% -------------------------------------------------------------
\subsection{Interpreting and Representing Objects over Time}
\label{sec:challenges:objects_over_time}
%%% -------------------------------------------------------------

%Integrating events, describing multiple observations in time, with an object-relation model in the same data model gives rise to a number of additional requirements and constraints for representing objects and relations. These seemingly invalidate a number of traditional, intuitive interpretations of objects and relations between objects.

An essential contribution of OCED is the ability to express how events relate to and operate on objects over time. If observed objects change over time, OCED needs a concept to describe \emph{changes to objects over time} and how different observations of the same object are related to each other. For instance, in Fig.~\ref{fig:example:po_created} and Fig.~\ref{fig:example:po_released}, PO object \emph{ID\#4829} changes the value for attribute \emph{release status} from \emph{[]} to \emph{[X]}.

\subsubsection{Event attributes.} A basic possibility is to encode changes in object attribute value changes as attributes of the event performing the change, e.g., by including event attributes \emph{[release\_status\_old]} and \emph{[release\_status\_new]}. However, this encoding requires consistent interpretation of event attribute names and cannot reliably describe value changes of multiple objects observed by an event.

\subsubsection{Static objects and attribute values changes.} A more reliable  possibility is to interpret an \texttt{object} as a static description of the entire object (e.g., its full representation at the moment of extraction) while historic changes to the object are represented elsewhere, e.g., in a series of timestamped attribute values that is proposed as an extension to OCED in OCEL2.0 and further discussed in Sect.~\ref{sec:challenges:strict_extensions}).

\subsubsection{Object snapshots.} Alternatively, one could interpret the concepts in \textcolor{blue}{Group B} of Fig.~\ref{fig:oced-mm:core} not as a static, singular observation of objects, attributes, and relations, but allow the same object $o$ to be observed repeatedly (in different states). In this interpretation, the same tangible or abstract entity $o$ can be observed multiple times. Each such observation of $o$ is represented as an \texttt{object} that is interpreted as a ``snapshot'' of $o$. The \texttt{observes} relation then describes that an event \texttt{observes} a specific ``snapshot'' of $o$. This idea extends to the attributes and relations associated with $o$ alike. 

\begin{figure}
    \centering
    \includegraphics[width=.85\linewidth]{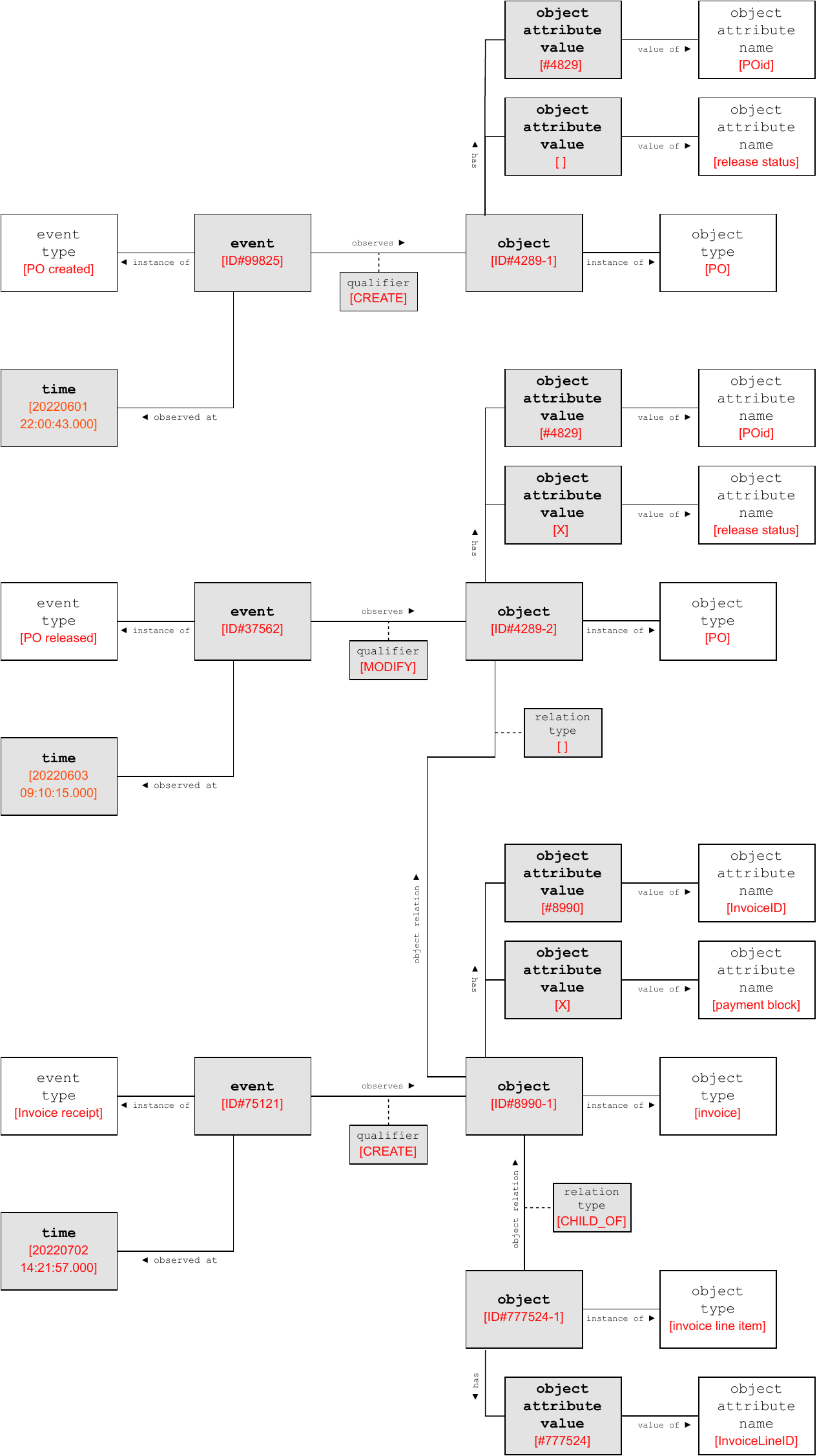}
    \caption{Snapshot interpretation of the example of Sect.~\ref{sec:oced_core:example}}
    \label{fig:example:snapshot}
\end{figure}

Figure~\ref{fig:example:snapshot} illustrates such a ``snapshot''-based representation of the events and objects of Figures~\ref{fig:example:po_created}, \ref{fig:example:po_released}, and \ref{fig:example:invoice_receipt} of the example in Sect.~\ref{sec:oced_core:example} together.

If such an interpretation is assumed, the following considerations arise:
\begin{itemize}
    \item Multiple ``snapshots'' of the same object $o$ must be relatable to each other. For instance, through an immutable \texttt{object attribute} that uniquely identifies the object (e.g., \emph{[POid]}, \emph{[InvoiceID]}, and \emph{[InvoiceLineID]} in Fig.~\ref{fig:example:snapshot}) and all ``snapshots'' of $o$ carry the same value (e.g., \emph{[POId=\#4829]}). 
    \item At the same time, object identifiers no longer identify objects but snapshots (e.g., \emph{\#4289-1} and \emph{\#4289-2} are different snapshots of the same \emph{PO} object).
    \item The moments in time where an object ``snapshot'' has been observed is described through the events that observe the snapshot; the same snapshot can be observed an arbitrary number of times, i.e., referred to from an arbitrary number of events.
    \item An object ``snapshot'' is not required to enumerate all object attributes but only those that are relevant for the observation or event, i.e., typically the changed attribute values.
    \item An object ``snapshot'' observed by an event itself may be ``empty'', i.e., it only contains the object identifier but no other object attributes etc. For example, when an event \texttt{observes} an object but does not change any of its attribute values (e.g., the event of a user loading a website). 
    \item Conversely, each object ``snapshot'' theoretically could be a complete representation of the entire object every time it is observed. While the OCED concepts do not dictate otherwise, common sense will render this approach non-viable given the storage footprint.
    \item This interpretation also includes (and hence generalizes) the minimal interpretation of OCED in Sect.~\ref{sec:oced_core:interpretation}: there is one ``full snapshot'' for each object that events can observe.
\end{itemize}
With these considerations in mind, Figure~\ref{fig:example:snapshot} describes the creation of a \emph{PO} object with \emph{[POid=\#4829]} and attribute \emph{[release status = []]} at time \emph{2022-06-01 22:00:43.000}; the subsequent \emph{PO release} event changes the \emph{PO}'s attribute to \emph{[release status = X]} at time \emph{2022-06-03 09:10:15.000}. The later \emph{Invoice receipt} event at time \emph{2022-07-02 14:21:57.000} creates an \emph{Invoice} object with \emph{[InvoiceID = \#8990]} and \emph{[payment block = X]} having an \emph{Invoice Line Item} with \emph{InvoiceLineID = \#777524}. At this time (\emph{2022-07-02 14:21:57.000}), \emph{Invoice} \emph{\#8990} is related to \emph{PO} \emph{\#4829} (in its most recently observed state).

\subsubsection*{Conclusion.} The OCED-MM Core Model described in Sect.~\ref{sec:oced_core} does not prescribe how to interpret an \texttt{object} allowing to use it also to describe auxiliary and proxy objects as discussed in Sect.~\ref{sec:challenges:relations_as_objects}, \ref{sec:challenges:attributes_as_objects}, \ref{sec:challenges:other_as_objects}. Thus, a snapshot interpretation of an object is possible. However, the baseline interpretation of an \texttt{object} describing \emph{the} object is arguably less ambiguous and complex in comparison.

Generally, describing changes to an object over time either requires an extension of the OCED model, i.e., timestamped attribute values, or a more involved interpretation of the \texttt{object} concept. Notably, different use cases may require different choices.

%%% -------------------------------------------------------------
\subsection{Semantics for Qualifiers and Relation Types}
\label{sec:challenges:qualifiers}
%%% -------------------------------------------------------------

OCED allows to \emph{qualify} the \texttt{observes} relation and giving a \emph{relation type} to an \texttt{object relation}. The purpose of defining qualifiers and relations is to provide semantics to events and object relations that allow consistent interpretation of object-centric event data across various tools and solutions. For this purpose, it is beneficial to agree on a set of qualifiers and relation types with agreed on semantics.

\subsubsection*{Agreeing on standard qualifiers and relation types.}
Suggested qualifiers that are likely to occur in most use cases are \texttt{CREATE}, \texttt{MODIFY}, and \texttt{DELETE}, while suggested object relation types that are likely to occur in most use cases are \texttt{CHILD\_OF} and \texttt{PARENT\_OF}). The following semantics have been proposed for these in the past.

\subsubsection*{Agreeing on standard qualifier semantics.}
Agreeing on qualifiers allows to agree on how to interpret how an event changes objects and associated relations. For instance, the semantics of a \texttt{CREATE} or \texttt{DELETE} qualifier on an \texttt{observes} relation between an \texttt{event} and an \texttt{object} implicitly extends to \texttt{object relations} the \texttt{object} can be interpreted as follows:
\begin{enumerate}
    \item Any \texttt{object relation}, if not created explicitly (i.e., after both objects are in existence), is created implicitly with the \texttt{CREATE} of the second \texttt{object}.
    \item Any \texttt{object relation}, if not deleted explicitly (i.e., while both objects are remaining in existence), is deleted implicitly with the \texttt{DELETE} of either \texttt{object}.
\end{enumerate}
For example, applying this interpretation on Fig.~\ref{fig:example:invoice_receipt} and Fig.~\ref{fig:example:snapshot} implies that the relation from \emph{Invoice} \emph{\#8990} to \emph{PO} \emph{\#4829} is implicitly created by event \emph{Invoice receipt} at time \emph{2022-07-02 14:21:57.000}. Agreeing on such an interpretation allows to omit further \texttt{observes} relations between the event and other objects.

\subsubsection*{Standard relation type semantics wrt.\ qualifiers.}
Likewise, agreeing on the semantics of relation types allows more extensive interpretations of events. For instance, \texttt{CHILD\_OF} defines an \texttt{object} relationship, in which an arbitrary number of child \texttt{objects} are related to exactly one \texttt{parent} object (N:1); the opposite applies to \texttt{PARENT\_OF}. The predefined object relation types (\texttt{CHILD\_OF}, \texttt{PARENT\_OF}) result in further implicit semantics between objects and events:
\begin{enumerate}
    \item When a child \texttt{object} gets \texttt{DELETE}d by an \texttt{event}, its \texttt{object relation} to the parent \texttt{object} is deleted in unison, however, the parent \texttt{object} itself is not affected.
    \item When a child \texttt{object} gets \texttt{CREATE}d by an \texttt{event}, its \texttt{object relation} to the parent \texttt{object} is created in unison, however, the parent \texttt{object} itself is not affected.
    \item When a parent \texttt{object} gets \texttt{DELETE}d by an \texttt{event}, all its child \texttt{objects} and their \texttt{object relations} to the parent are deleted in unison.
    \item When a parent \texttt{object} gets \texttt{CREATE}d, its initial child \texttt{objects} (i.e., the ones without a dedicated related \texttt{CREATE} event) are created in unison (e.g., when a \emph{purchase order} is created with ten \emph{line items}, solely the \emph{purchase order}’s link to the event needs to be captured).
\end{enumerate}
For example, applying this interpretation on Fig.~\ref{fig:example:invoice_receipt} and Fig.~\ref{fig:example:snapshot} implies that \emph{Invoice Line Item} \emph{\#777524} is implicitly created as a \emph{CHILD\_OF} \emph{Invoice} \emph{\#8990} by event \emph{Invoice receipt} at time \emph{2022-07-02 14:21:57.000}.

Assigning such semantics allows to omit certain relations in the data, especially the \texttt{observes} relations from events to a large set of objects, reducing the data footprint for transmission and storage, while allowing the omitted relations (and event semantics) to be unambiguously reconstructed.

%%% -------------------------------------------------------------
\subsection{Relations between Objects}
\label{sec:challenges:relations}
%%% -------------------------------------------------------------

In the core model, the \texttt{object relation} from \texttt{object} to \texttt{object} only models the \emph{existence} of a relation of a particular type between two objects. The relation itself does not bear\footnote{While strictly speaking the relation may have a technical identifier, the relation is already fully identified by the pair of source and target node. This is due to a non-trivial and relevant detail of conceptual modeling: a conceptual model as in Fig.~\ref{fig:oced-mm:core} describes a family of graphs. Concepts describe the allowed nodes, relations describe the allowed edges that may be present between two nodes in this graph, i.e., pairs of nodes. As such there cannot be two distinct edges between the same pair of nodes.} any identifier and thus cannot be referred to. Any \texttt{object relation} is implicitly and uniquely identified by the pair (source \texttt{object}, target \texttt{object}) while \texttt{relation type} is only a \emph{qualifier} for the relation, i.e., describes the nature of the pair (source \texttt{object}, target \texttt{object}).

A strict interpretation of this concept allows that two objects are related by at most one kind of relation, but two objects may not be related in two different ways. For example, assume \emph{[Order o]} is owned by \emph{[Person p]}, and also \emph{[Order o]} has been issued by the same \emph{[Person p]}. The OCED-MM Core Model can only represent the binary relation $(p,o)$ \,--\,which can exist only once\,--\,and qualifies it, either by ``owned by'' or by ``issued by''. Most source systems do not have this restriction and identify a relation by the triple (source \texttt{object}, target \texttt{object}, \texttt{relation type}). This challenge has to be addressed through conventions or extensions of OCED, e.g., materializing relations as objects (see Sect.~\ref{sec:challenges:relations_as_objects}).

Independently of this uniqueness constraint, further aspects of relations have to be considered. 
While the OCED meta-model shall be decoupled from any particular storage format (see Sect.~\ref{sec:path_to_oced}), general constraints of serializing OCED in a representation for storage to data exchange need to be considered. Specifically wrt.\ relations, implementations of OCED have to be aware of the following considerations.

\subsubsection{Serializing References.} When serializing any actual instance of OCED, an \texttt{object relation} can be serialized as follows:
\begin{itemize}
    \item As an explicit triple (source, target, type). However, a relation in such a representation is inherently ``static'', i.e., it denotes that such a relation exists between the source and target \texttt{object} but no event can refer to such a tuple to express changes to the relation. Sect.~\ref{sec:oced_core:concepts} discusses in which cases the semantics of observing \texttt{CREATE} or \texttt{DELETE} of an object implies creation and deletion of associated \texttt{object relations}. But OCED cannot express, for instance, changes to relations due to \texttt{MODIFY} events (e.g., reassigning a \emph{[package]} object from one \emph{[delivery]} object to another \emph{[delivery]} object). Expressing such dynamics requires to ``objectify'' the relation, see Sect.~\ref{sec:challenges:relations_as_objects}.

    \item As an implicit triple stored as attribute at one of the objects, i.e., similar to a foreign-key attribute, an \texttt{object attribute} (both, name and value) is explicitly marked as ``reference attribute'' of the source object having as
        \begin{itemize}
            \item \texttt{object attribute name} the relation type
            \item \texttt{object attribute value} the identifier of the target \texttt{object} (see Sect.~\ref{sec:oced_core:interpretation}).
        \end{itemize}
    Thereby, usual principles of data serialization apply. 
    \begin{itemize}
        \item Serializing a 1:N relation, requires the objects at the N-side to store the reference attribute to the object at the 1-side\footnote{Storing a reference to N objects at the 1-side would violate the requirement that attribute values do not have complex structures in themselves}.
        \item An N:M relation cannot be serialized in this way, but requires either storage of all relations as triples or reification of the relation into an object (Sect.~\ref{sec:challenges:relations_as_objects}).
    \end{itemize}
\end{itemize}

\subsubsection{Reference Attributes vs Relations: Source of Potential Inconsistencies.} The alternative forms for serializing object relations described above is a source of potential inconsistencies in OCED instances that any implementation of OCED (both producers and consumers of OCED) have to address:
\begin{enumerate}
    \item Source data may contain object attributes which explicitly or implicitly refer to other objects, i.e., model relations. In OCED these should either by fully expressed and serialized in the way object relations are serialized, or the object reference attribute and the relation must state the same values (i.e., there are no two conflicting descriptions of a relation).
    \item Attributes not marked as relation references must not be interpreted as relation references.
\end{enumerate}

\subsubsection{Events referring to multiple objects.} \label{subsubsec:events_referring_to_multiple_objects}
Also events can be a source of inconsistency in relations. Events referring to multiple objects may (implicitly) suggest relations between these objects that are not explicitly represented in the data. Consider the following example: ``Event \emph{[\#e17 Clear Invoice]} observes object \emph{[Payment P1]} and objects \emph{[Invoice I1]} and \emph{[Invoice I2]}'', i.e., payment P1 is used to clear Invoices I1 and I2. Depending on the semantics of the event and the involved objects, this may constitute a relation between P1 and I1 and between P1 and I2. The current OCED-MM Core Model does not state whether these relations also have to be described and whether any integrity constraints between object relations and objects observed together at events apply. 

Future developments of OCED must provide clear conventions to producers and consumers of OCED  wrt. implicit and explicitly represented relations.

%%% -------------------------------------------------------------
\subsection{Relations as Objects}
\label{sec:challenges:relations_as_objects}
%%% -------------------------------------------------------------

In the OCED-MM Core Model, only objects and events are required to be uniquely identified, allowing to describe how objects change over time (see Sect.~\ref{sec:oced_core:interpretation} for the associated interpretations and design decisions and Sect.~\ref{sec:challenges:objects_over_time} for describing changes of objects over time). 

\subsubsection{Changes to object relations.}
As object relations do not carry identifiers, the OCED-MM Core Model is limited wrt. describing how object relations change over time.

For example, suppose an OCED instance records 
\begin{itemize}
\item objects \emph{[student S]}, \emph{[supervisor M]}, and \emph{[supervisor D]},
\item an event $e1$ observing \emph{[student S]} with a \emph{[supervises]} relation to \emph{[supervisor M]}, and 
\item a second event $e2$ observing \emph{[student S]} with a \emph{[supervises]} relation to \emph{[supervisor D]}.
\end{itemize}
It is not possible to conclude whether the supervisor of student $S$ changed from $M$ to $D$, and that the prior relation from $S$ to $M$ no longer holds. It may be equally possible that $M$ and $D$ both supervise $S$ but no single event observes this, or that $M$ and $D$ alternate supervision.

Generally, an event $e$ observing an object $o1$ with a relation $R$ to an object $o2$ only states that, at the moment of $e$, the relation $R$ from $o1$ to $o2$ has been ``observed''. No further interpretation is possible: Observing the relation $R$ from $o1$ to $o2$ by event $e$ for the first time does not imply that it has been created at that point. Neither does observing it by event $e$ for the last time imply that is has been deleted. 

\subsubsection{Usage pattern or explicit extension.}
Reliably and unambiguously tracking object relation across multiple observations in time requires to assign identifiers to object relations. This was proposed in the OCED-MM Full Model (see Sect.~\ref{app:oced-mm-original}) but can also be achieved within the OCED-MM Core Model by materializing (also called \emph{reifying}) an object relation \emph{(source, target, T)} into an object on its own. In other words, by introducing an ``artificial'' object $R$ of type $T$  that has a \emph{from} and a \emph{to} relation to the \emph{source} and \emph{target} objects, see Fig.~\ref{fig:oced-mm:full}. This artificial relation object now carries an identifier and can be explicitly observed by events through qualified relations.

Note that materializing a relation is a potential source of inconsistencies and ambiguities in OCED, see Sect.~\ref{sec:challenges:cycles_ambiguity}.

%%% -------------------------------------------------------------
\subsection{Attributes as Objects}
\label{sec:challenges:attributes_as_objects}
%%% -------------------------------------------------------------

Object attributes are limited to basic types and are not required to carry an identifier.

\subsubsection{Complex data types.}
Use cases that require to represent complex data types can do this by introducing a proxy object which in turn can have a collection of object attributes (or refer to further objects). For example, an attribute representing a numeric value and a unit can be stored as a proxy object containing solely the numeric value and the unit (as text) and then be related to the original object. Alternatively, such compound values can be represented through a collection of \texttt{object attribute}s with consistent naming conventions, e.g., \emph{price} and \emph{price\_unit}. 

\subsubsection{Object attribute changes.}
Some use cases require expressing that an event operated on a specific \texttt{object attribute} or in which way.

In some cases, naming conventions for \texttt{object attribute}s can help express the semantics of an event with respect to object attributes. For instance, if an event \emph{[Price Change]} changed attribute \emph{[price]} of object \emph{[Order\#17]}, then naming conventions for \texttt{event attribute}s can help expressing the semantics of an event without the need for creating proxy objects, e.g., \emph{[price\_old]} and \emph{[price\_new]}. 

However, it may be inconvenient to represent a series of multiple attribute-value changes of an object in this way as one rather wants to express that the event directly (and only) operated on the particular attribute. This either requires extending OCED, see (Sect.~\ref{sec:challenges:strict_extensions}) or materializing attributes as objects. That is, turn a particular attribute of an object $o$ into a uniquely identifiable object $A$ that in itself carries the value as an attribute-value pair. Then, events can explicitly refer to $A$ through different qualified relations.

\subsubsection{Fine-grained semantics.}
Also, there may be use cases requiring to express event semantics on a more fine-grained level, e.g., when (as a result of more complex database transaction) an event $e$ is, both, \texttt{READ}-ing an attribute $a$ of an object $o$, \texttt{DELETE}-ing $o$, and \texttt{CREATE}-ing another object $o2$ whose values depend on $a$. Also, in this case, materializing the attribute as an object $A$ would allow to express the semantics of $e$.

Note that materializing an object attribute is a potential source of inconsistencies and ambiguities in OCED, see Sect.~\ref{sec:challenges:cycles_ambiguity}.

%%% -------------------------------------------------------------
\subsection{Other Process Concepts as Objects}
\label{sec:challenges:other_as_objects}
%%% -------------------------------------------------------------

Further limitations of the OCED-MM Core Model (listed in Sect.~\ref{sec:oced_core:limitations}) can be addressed through introducing artificial objects.

Events may be indirectly related through shared (abstract) objects, however such semantics are not prescribed by the meta-model. Examples: 
\begin{itemize}
    \item duration of an activity can be achieved through an (abstract) proxy object linking to exactly two events: start and end, 
    \item transaction types can be represented through an (abstract) proxy object linking to multiple events (e.g., [scheduled], [started], [completed], [archived]), or
    \item grouping of events can be achieved through an (abstract) proxy object linking to multiple events (i.e., groups of events).
\end{itemize}

The limitation of non-complex data types can be overcome by defining (abstract) proxy objects that are related to the main object via an object relation (e.g., CHILD\_OF). With object attribute values of the child objects, nesting can be mimicked. 

While the OCED-MM Core Model supports the creation of artificial objects and proxy objects to describe more complex structures in event data and objects, clear conventions must be established to ensure interoperability between OCED producers and consumers. For instance, in case a proxy object $A$ relates multiple events $e_1,\ldots,e_k$ to each other to describe a long running activity operating on several objects $o_1,\ldots,o_l$, can each event $e_i$ refer to any object $o_j$, or do all events refer to all objects, or may only $e_1$ or $e_k$ refer them, etc.?

%%% -------------------------------------------------------------
\subsection{Strict OCED extensions}
\label{sec:challenges:strict_extensions}
%%% -------------------------------------------------------------

\subsubsection{Timestamped Attribute Values.}

Events are considered atomic and have a timestamp. Therefore, event attributes are fixed.
Objects may evolve over time and be involved in multiple events. Therefore, object attributes may change as illustrated in Sect.~\ref{sec:oced_core:example} in Fig.~\ref{fig:example:po_created} and Fig.~\ref{fig:example:po_released}. The baseline interpretation of the OCED-MM Core Model cannot describe such changes as explained in Sect.~\ref{sec:challenges:objects_over_time}

One possible way to handle this is to extend object attributes with a timestamp.
OCEL 2.0 supports so-called \textit{dynamic attribute values} that can change over time \cite{ocel2_specification,DBLP:conf/icpm/KorenABA23}. 
Instead of having a single, fixed value, 
an object attribute may have a value that changes over time.
The smallest timestamp in OCEL 2.0 is 1970-01-01 00:00 UTC. This is the default time of an attribute value. 
Later values for the same attribute should be seen as updates.
For example, if an object attribute \textit{weight} has a value of 80 kilograms with timestamp 2023-01-01 00:00 UTC,
a value of 90 kilograms with timestamp 2024-01-01 00:00 UTC, 
and a value of 85 kilograms with timestamp 2025-01-01 00:00 UTC,
then the weight of the object is assumed to be 80 kilograms throughout the year 2023 and 90 kilograms throughout the year 2024.
The timestamps may coincide with the timestamps of events, allowing for some form of event correlation. 
However, this is not mandatory.

\begin{figure}
    \centering
    \includegraphics[width=\linewidth]{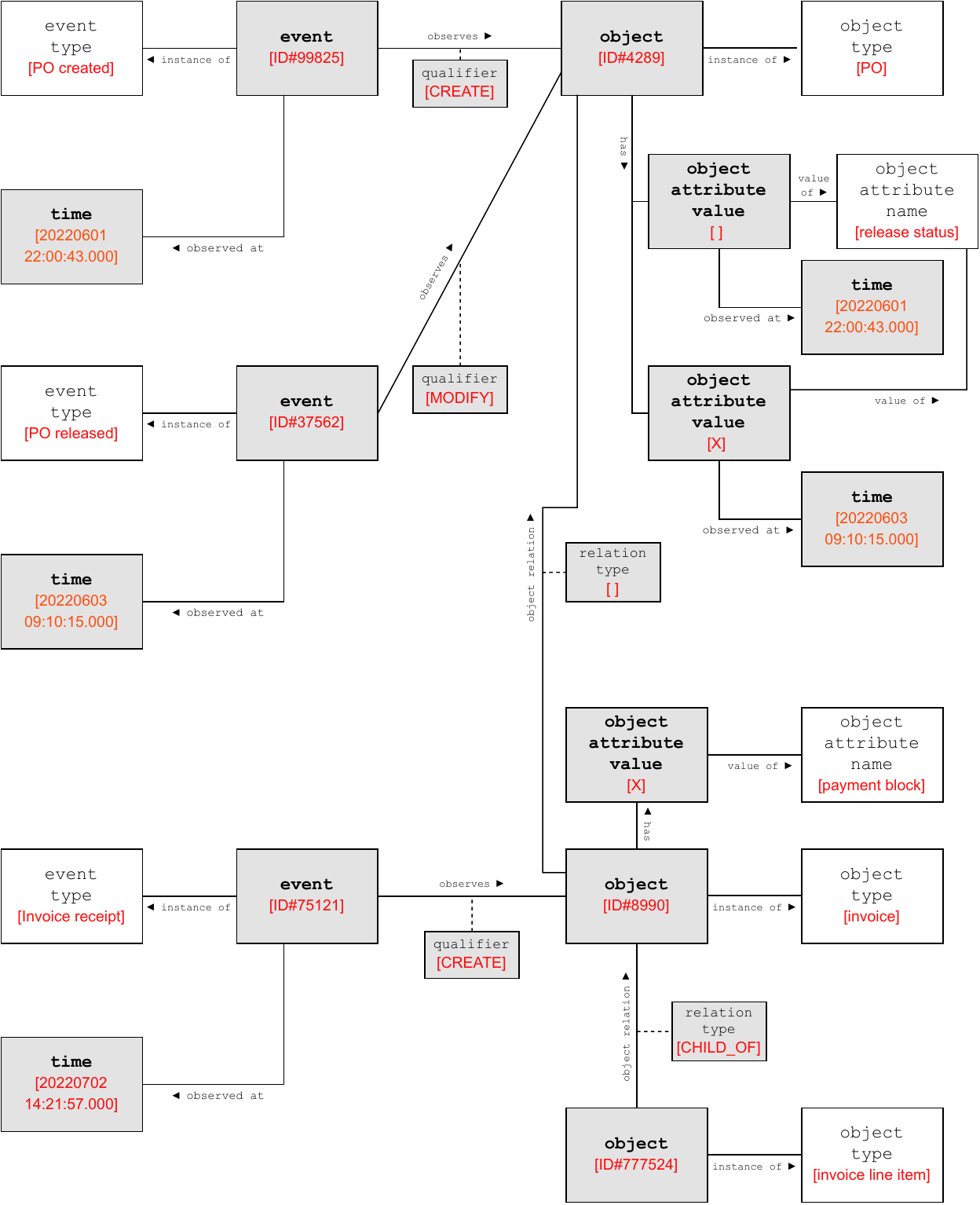}
    \caption{Running Example with OCEL 2.0 timestamped attribute values}
    \label{fig:example:ocel2}
\end{figure}

Figure~\ref{fig:example:ocel2} illustrates this extension by extending the baseline interpretation of the running example in Fig.~\ref{fig:example:minimal} with timestamped attributes values: the \emph{PO} object now \texttt{has} \emph{two} \texttt{object attribute values} for the \texttt{object attribute} \emph{release status}. Value \emph{[]} has been \texttt{observed at} time \emph{2002-06-01 22:00:43.000} while value \emph{[X]} has been \texttt{observed at} time \emph{2002-06-03 09:10:15.000}. Together, they describe that and when the attribute value changed, which is not described in the baseline interpretation in Fig.~\ref{fig:example:minimal}. Note that while these timestamps coincide with timestamps of the event creating and releasing the \emph{PO}, and thereby implicitly are correlated with them, this correlation is not mandatory.

\subsubsection{Data Schemas.} The OCED-MM Core Model describes the existence of type information for events and event attributes, and for objects and object attributes, but does not describe relations between \texttt{event type} and \texttt{event attribute} and \texttt{object type} and \texttt{object attribute}. It also does not allow to express which types of \texttt{object relations} may exist between which \texttt{object types}. Such information, typically documented in a \emph{data schema}, informs tools about which attributes are reliably present in all \texttt{instances of} a particular event or object, allowing algorithms and statistics to draw on them. Further, object, events, and relations may in practice carry \emph{more than one type}, for instance when expressing generalization and specialization or when describing different functions of them (e.g., a specific employee can be a \emph{resource} as well as a \emph{role}).

Data schemas cannot be expressed through (repurposing) existing concepts of the OCED-MM Core Model and requires a true extension. Two such extensions have been explored in OCED implementations (see Sect.~\ref{sec:implementations}).
\begin{itemize}
\item OCEL 2.0 allows for the specification of the possible \texttt{event attributes} per \texttt{event type} and the possible \texttt{object attributes} per \texttt{object type}. This is also reflected in the OCEL 2.0 meta-model.
\item The \emph{OCED-PG} implementation~\cite{DBLP:conf/icpm/SwevelsFM23_oced-pg} (cf., Sect.~\ref{sec:implementation:oced-pg}) of OCED argues that any instance of OCED has to be accompanied by an OCED-compliant domain-specific schema of the data. This schema should specify the specific types of objects, relations, and events present in the data and how they are related to each other. Similar to \emph{OCEL 2.0}, \emph{OCED-PG} extends the OCED Meta-Model to also specify attributes of objects and events, but also includes relations. The extended OCED Meta-Model can then be refined into a domain-specific OCED-compliant schema. But current implementations do not support checking consistency of an OCED instance to a given schema.
\end{itemize}

%%% -------------------------------------------------------------
\subsection{Cycles and Ambiguity through Extensions}
\label{sec:challenges:cycles_ambiguity}
%%% -------------------------------------------------------------

Materializing relations or attributes in artificial objects introduces \emph{cycles} in the data model that involve the \texttt{observed} relation between events and objects. Also other extensions, such as assigning timestamps to other concepts introduce cycles. These cycles are a possible source of inconsistency in representing and interpreting OCED.

\subsubsection{Overloading the object concept.} The preceding sections presented various use cases for a flexible interpretation of the \texttt{object} concept besides the tangible and abstract entities present in the process itself. These interpretations of objects include ``snapshots'' (Sect.~\ref{sec:challenges:objects_over_time}), relations (Sect.~\ref{sec:challenges:relations_as_objects}), attributes (Sect.~\ref{sec:challenges:attributes_as_objects}), and other process concepts such as activities or transactions (Sect.~\ref{sec:challenges:other_as_objects}). Allowing these interpretations overloads the \texttt{object} concept itself. Consistent interpretation across producers and consumers requires conventions and mechanisms to clearly distinguish what exactly an object stands for.

\subsubsection{Materializing relations as objects.} When materializing an \texttt{object relation} as an object $R$, the attributes and relations of $R$ have to be consistent with all other representations of this relation in the data which includes (changes to) reference attributes describing the relation (see Sect.~\ref{sec:challenges:relations}). For example, assume an object \emph{[package \#5]} has an \texttt{object attribute} \emph{[assigned-to] = [delivery tour \#17]} which is interpreted as a relation \emph{([package \#5],[delivery tour \#17],[assigned-to])}. To model changes to this relation, a relation object $R$ is created representing this relation \emph{([package \#5],[delivery tour \#17],[assigned-to])}. Subsequently, an event $e$ changes the delivery tour assigned to an object \emph{[package \#5]} by updating $R$ into \emph{([package \#5],[delivery tour \#23],[assigned-to])}. An event $e$ \texttt{observing} $R$ has to be consistent with observations of all involved objects and their attributes along these changes. Note that this can be expressed in OCED, both, through the ``snapshot'' interpretation of objects (Sect.~\ref{sec:challenges:objects_over_time}) as well as through timestamped attribute values of reference attributes (Sect.~\ref{sec:challenges:strict_extensions} and Sect.~\ref{sec:challenges:relations}), and hence consistency considerations arise either way.

\subsubsection{Materializing attributes as objects.} For example, assume an event $e$ observes object \emph{[Order 17]} which is related to a materialized attributed object \emph{[Order 17+price]}. Should $e$ also have an \texttt{observes} relation to \emph{[Order 17+price]} or is this \texttt{observes} relation implicit? Likewise, if $e$ observes, both, \emph{[Order 17]} and \emph{[Order 17+price]}, do the qualifiers of both relations have to be identical or can they differ, e.g., \texttt{CREATE} \emph{[Order 17]} and \texttt{MODIFY} \emph{[Order 17+price]}. Extracting OCED from source systems may result in such cases.

\subsubsection{Assigning timestamps to other concepts.} OCEL 2.0 proposes to extend the OCED-MM Core Model to let an \texttt{object attribute value} $v$ have a \texttt{timestamp} $t$. While this allows for an efficient representation (and extraction) of a series of changes to an \texttt{object attribute} $a$, it introduces an new, implicit representation of an \texttt{event} $e$ that \texttt{observes} an \texttt{object attribute value} $v$ for attribute $a$ at time $t$. 

Consumers of OCED with such an extension must be aware of the semantics of this extension and must agree on whether the implicit representation of an event is to be converted into an explicit representation of the event and which conventions to follow. For instance, the explicit representation of the event may require to materialize an object attribute as an artificial object itself (with all associated considerations discussed above). Further, as illustrated also in Fig.~\ref{fig:example:ocel2}, timestamps of \texttt{object attribute values} may coincide with timestamps of events that \texttt{observe} the owning \texttt{object} suggesting event correlation, e.g., that the object attribute value \emph{[]} has been set by the \emph{PO created} event that has been observed at the same time. OCED analysis techniques have to be aware of and identify such correlations, and more involved consolidations of the implicitly and explicitly represented events are required to obtain an unambiguous representation.

%%% =============================================================
\section{Initial Implementations and Lessons Learned}
\label{sec:implementations}
%%% =============================================================

Following the ``Call for Reference Implementations'' for the OCED Meta-Model proposal, four independent implementations were presented and discussed at the OCED Symposium of the 5th International Conference on Process Mining (Rome, 2023) while a fifth implementation was recently published.

%%% -------------------------------------------------------------
\subsection{Konekti}\label{sec:implementation:konekti}
%%% -------------------------------------------------------------
\href{www.getkonekti.io}{Konekti} is a commercial data transformation tool for process mining. It reduces the required effort for process mining, allowing practitioners to focus on higher-value tasks. Its primary focus is on document-based information systems, such as ERP, CRM, and WMS platforms. Through our industry experience, we have identified that the most efficient method for generating case-centric event logs from document-based systems is by first constructing an object-centric data model. This model can serve as a staging layer from which to create case-centric logs or be used for object-centric analysis. Hence, Konekti supports both object-centric and case-centric analysis.  

\subsubsection*{Key Features of Konekti.}
Konekti offers several key features that distinguish it from other process mining tools:
\begin{itemize}
    \item \textbf{Guided object-centric meta-modeling} ensures that all users follow a consistent meta-model, promoting collaboration and knowledge transfer. Konekti also suggests next steps to streamline workflow.
    \item \textbf{Built-in data quality checks} accelerate the data validation process by automatically detecting potential issues.
    \item \textbf{Conversion to case-centric logs} enables users to convert the object-centric data model into case-centric logs for use in conventional process mining analysis tools.
    \item \textbf{Automated script generation} reduces the effort and skill required. Users follow a low-code workflow to construct the data model, after which Konekti generates an exportable PostgreSQL or Spark script.
    \item \textbf{Interactive object-centric data model visualization} improves comprehension of the data model and fosters collaboration among users.
\end{itemize}

\subsubsection*{OCED-MM Implementation in Konekti.}
To expedite the production of case-centric event logs, certain elements from the OCED-MM model are omitted in Konekti’s implementation:
\begin{enumerate}
    \item \textbf{Object-relation types:} Konekti does not implement predefined relation types (such as \texttt{CHILD\_OF} or \texttt{PARENT\_OF}) as outlined in Section \ref{subsubsection:object_relation}. Instead, relationships are inferred implicitly through their cardinality (1:1, 1:n, n:m), which is computed by Konekti.
    \item \textbf{Event-to-object relation qualifiers:} Konekti omits the predefined qualifiers (\texttt{CREATE}, \texttt{MODIFY}, \texttt{DELETE}) described in Section \ref{subsubsection:observes_relation_from_events_to_objects} and \ref{sec:challenges:qualifiers}. Instead, event qualifiers are embedded in the activity name (e.g., “create,” “modify,” or “delete”), and Konekti computes the cardinality of event-to-object relations (1:1, 1:n, n:m).
\end{enumerate}
As a minor extension, event types in Konekti can have zero events instantiated, deviating from the event definition in Section \ref{subsubsec:event}.

\subsubsection*{Lessons Learned.}
Our experience of implementing Konekti has led to several insights regarding object-centric data modeling. Below, we list our two key insights:
\begin{enumerate}
    \item \textbf{Relating event types to a single object type minimizes complexity.} 
    Ambiguity often arises regarding which objects should be related to an event. Although Konekti allows linking an event to multiple object types, we recommend associating each event type with a single object type. For example, in this example introduced in Section \ref{subsubsec:events_referring_to_multiple_objects}:  
    \begin{quote}
        ``Event \#e17 [Clear Invoice] observes object Payment P1 and objects [Invoice I1] and [Invoice I2], where payment P1 is used to clear invoices I1 and I2.''
    \end{quote}
    We advise to initially link Event \#e17 only to Payment P1 and model the relationship between Payment P1 and the two invoices (I1 and I2) through an object relation. If required for object-centric analysis, users can later add additional event-object relations. This way of working reduces the model complexity.
    \item \textbf{Ambiguity in the definition of objects.} In Section \ref{subsubsection:object}, an object is defined as a uniquely identifiable entity observed in the world. However, this definition can be blurred due to several factors:
    \begin{enumerate}
        \item \textbf{Objects as attributes:} Not every identifiable entity needs to be modeled as an object. For instance, while ``Customer'' could be an object, it might also be modeled as attributes (e.g., CustomerId, CustomerName) of the ``Order'' object. The decision on whether to treat an entity as an object or an attribute is context-dependent and determined by the data modeler.
        \item \textbf{Hierarchical structures in objects:} The level of granularity for modeling objects can vary. A user could model one object ``Financial document'' or decide to split them into multiple objects (e.g., ``Invoices'', ``Credit memos'', etc.). The best choice depends on the context, making it difficult to establish a universal best practice.
        \item \textbf{Attributes as proxy objects:} In some cases, attributes can be modeled as objects. For example, while ``Order Status'' might seem better suited as an attribute, it can be modeled as a separate object, aligning with systems like SAP ECC, where status information is stored in separate tables (e.g., table VBUK for sales document statuses).
        \item \textbf{Relations as proxy objects:} Konekti expects relationships between objects to be modeled through object tables (using primary key-foreign key combinations). If relations are stored in separate tables, this relation needs to be modeled as an object (e.g., ``Sales Document Flow''), even though such tables are not real-world entities.
        \item \textbf{Grouping activities via proxy objects:} As described in Section \ref{sec:challenges:other_as_objects}, proxy objects can be used to group activities logically within the data model.
    \end{enumerate}
    The inherent ambiguity in defining objects contributes to variability in how process data models are constructed, potentially complicating the understanding, modeling, and validation processes.
\end{enumerate}

\subsubsection*{Conclusion.}
In short, Konekti streamlines the process of generating object-centric and case-centric event logs using an OCED-MM implementation. While challenges remain in consistently relating events to objects and defining objects clearly, our experience with Konekti demonstrates that the model is effective and practical in real-world applications.

%%% -------------------------------------------------------------
\subsection{OpenOCED}\label{sec:implementation:open_oced}
%%% -------------------------------------------------------------

%Scope of implementation (wrt. OCED meta-model (core model in Sect. 3, base/full model in App A)

\emph{OpenOCED} is an open source reference implementation using a Model Driven Engineering (MDE) perspective of the \emph{OCED-MM Core Model} of Sect.~\ref{sec:oced_core}. The MDE approach and first Java library implementation of the OCED base proposal from March 2023 by the IEEE Task Force on Process Mining was presented in \cite{DBLP:conf/bpm/Calegari023}. To represent OCED models compliant with the Ecore meta-model the XML Metadata Interchange (XMI) standard ({\url{https://www.omg.org/spec/XMI/}) format is used, wich can be trasnformed to JSON and XML OCEL 2.0 files, as well as CSV files. The \emph{OCED-MM Full Model} of App.~\ref{app:oced-mm-original} is also defined as an extension of the base meta-model using the EMF standard extension mechanism. 

%Functionality and Features
\subsubsection*{Features.}
The OpenOCED reference implementation provides a Python and a Java library that includes:
\begin{itemize}
    \item Ecore-based definition of the OCED meta-model.
    \item Import/export from/to XMI files of OCED model instances.
    \item Import/export from/to CSV files of OCED model instances.
    \item OCED model transformation to OCEL 2.0 (JSON and XML format).
    \item OCEL 2.0 (JSON and XML format) model transformation to OCED.
\end{itemize}

\subsubsection*{OCED-MM implementation in OpenOCED.} The OpenOCED meta-model directly represents the original object-centric concepts and expresses relations through qualified associations represented as pivot elements (e.g., event-object, object-object) to store its qualifier. The implementation does not enforce pre-defined qualifiers for the relations and allow navigation from the object-object relation to its source and target objects and from the object to the relations in which the object participates. Input files can be in XMI/CSV/OCEL 2.0 JSON and XML formats, the Java and Python code automatically generated from the meta-model allows in-memory manipulation of OCED models, which in turn can be output in XMI/CSV/OCEL 2.0 JSON and XML formats. Initial steps have also been taken to extend the interoperability with other OCED implementations, e.g., input/output to OCED-PG. 

OpenOCED libraries can be integrated into several Java or Python software to serve as an exchange format for existing Object-Centric Process Mining (OCPM) \cite{DBLP:conf/sefm/Aalst19} techniques and tools to support the manipulation of OCED models and/or to implement new algorithms for the OCPM perspectives. Also, with the MDE meta-model extension mechanism, the OCED-MM Core Model and Full Model can be extended to support other approaches directly, for which the associated code can be automatically generated. 

%Lessons Learned (and further developments)
\subsubsection*{Lessons Learned.} As lessons learned, a key aspect to consider when manipulating OCED models is information loss when extracting data and representing relationships between elements or exchanging different representations/implementations based on mappings. Also, to better exploit object relationships, e.g., in the context of data refinement (e.g., domain-specific filters applied to the model) and conformance checking (e.g., adding domain rules), the provision or extension of tools is needed. Current and future work includes applying OpenOCED to case studies with actual data to further deal with these key aspects, and extend the capabilities of the libraries to support more functionalities.

%Links to tool/resources
\subsubsection*{Resources.} Open OCED code and libraries for Python and Java are available at (\url{https://open-coal.pages.fing.edu.uy/oced/}), as well as the meta-model (Ecore, PyEcore) definition and examples in .XMI, CSV, and OCEL 2.0 JSON and XML format from (\url{https://ocel-standard.org/event-logs/overview/}).

%%% -------------------------------------------------------------
\subsection{Event Knowledge Graphs (OCED-PG)}\label{sec:implementation:oced-pg}
%%% -------------------------------------------------------------

%Scope of implementation (wrt. OCED meta-model (core model in Sect. 3, base/full model in App A)
\emph{OCED-PG}~\cite{DBLP:conf/icpm/SwevelsFM23_oced-pg} is an open-source library that uses Labeled Property Graphs (LPGs) to represent and store OCED in a (Neo4j) graph database. OCED-PG specifically addresses the use case of transforming (arbitrary) source data into an OCED store without an intermediate OCED-specific data exchange format, and querying OCED using mature graph query languages, enabling object-centric process mining using graph databases.

\subsubsection*{Features.}
OCED-PG formalizes the OCED meta-model using \emph{PG-schema}, the schema definition language for property graphs. Thus, the PG-schema formalization of OCED defines the concepts for storing and querying OCED in a graph database, enabling the following features:
\begin{itemize}
    \item \textbf{Refining the OCED data schema into a domain-specific OCED data schema}, which allows to explicitly document the object types, relation types, and event types present in the data.
    \item \textbf{Pattern-based transformation rules}, which allow describing a mapping from raw source data (in tabular format) into the domain-specific OCED data schema, the so-called \emph{semantic header}.
    \item \textbf{A fully automated pipeline} for loading and transforming raw source data into OCED.
    \item \textbf{Querying OCED using domain-specific concepts} by using the graph database's native query language over a domain-specific data schema.
\end{itemize}

\subsubsection*{OCED-MM implementation in OCED-PG.}
OCED-PG implements the \emph{OCED-MM Core Model} of Sect.~\ref{sec:oced_core} and the \emph{OCED-MM Full Model} of App.~\ref{app:oced-mm-original}. The OCED core model interpretation is inherent in the model of \emph{Event Knowledge Graphs} (EKG)~\cite{esser_2020_multidimensional} on which OCED-PG is based. Refining the EKG concepts allows to express OCED-MM Core Model extensions, which are currently implemented for the OCED-MM Full Model. The current semantic header allows creation of OCED-MM Full Model instances with some limitations: (1) All \texttt{observes} relations from one event receive the same qualifier (i.e., cannot distinguish different qualifiers originating from the same event). (2) Generating \texttt{observes} relations to \texttt{object attribute values} or \texttt{object relations} uses relation inference from existing relations (via the \texttt{object} associated to the attribute or relation). While more involved, the latter does ensure semantic consistency between relations, see Sect.~\ref{sec:challenges:cycles_ambiguity}.

\subsubsection*{Lessons Learned.}
OCED-PG is based on lessons learned in a series of industrial and academic case studies for extracting and modeling object-centric event data using graph databases and subsequent process mining analyses.
\begin{enumerate}
    \item \textbf{Domain-specific schemata:} Industrial use cases benefited from the development of a domain-specific data schema in which the OCED core concepts of events, objects, and their relations are refined into domain concepts (e.g., distinguishing object types, relation types, and groups of event types).~\cite{chu2022,cuprinsu2022,broniewski2023,DBLP:conf/bpm/SwevelsDF23,marangoz2023}

    \item \textbf{Structuring Schema-less Source Data:} Raw source data is often schema-less, or available schema information is not aware of events. Extracting OCED requires to first bring structure to the source data by specifying which attributes constitute information for which OCED concept, e.g., a record with a timestamp and an activity allows extracting an event.
    
    \item \textbf{Disentangling Events and Objects:} Source records are not inherently events or objects but may contain attributes describing multiple concepts (events, different objects and attributes). Extraction of OCED requires methods to disentangle them.

    \item \textbf{Some relationships are implicit:} Relationships can be established based on the co-occurrence of two elements (events and/or objects) within a record, using provenance as a basis (see Sect.~\ref{sec:challenges:relations}). Additionally, relationships can be inferred from existing relationships. For example, if a member has borrowed a book listed in a library catalog, it implies that the member is a member of the library.

    \item \textbf{Object and relation snapshots:} In industrial configuration management, source data includes snapshots of data objects and relations (describing the creation and evolution of configuration management data structures), requiring a ``snapshot'' interpretation of objects and relations (see Sect.~\ref{sec:challenges:objects_over_time}) and clear qualifier semantics (see Sect.~\ref{sec:challenges:qualifiers})\footnote{\url{https://www.linkedin.com/pulse/boost-your-knowledge-graph-events-gain-untapped-martijn-dullaart-wunse/}}. The results suggest it is beneficial to generalize timestamped attribute values of Sect.~\ref{sec:challenges:strict_extensions} to entire ``snapshots'', i.e., time-stamping objects and relations representing their state.~
    \cite{marangoz2023}
    
    %\item \textbf{Objects as event context:} Use cases for industrial processes in a physical context (e.g., in logistics and manufacturing) required to represent not only describe the objects handled by the process, but also objects describing the resources and systems in detail (e.g., manufacturing system components and layout) as context for events, enabling querying, inference, and various process mining use cases.~\cite{chu2022,cuprinsu2022,broniewski2023,DBLP:conf/bpm/SwevelsDF23}

    \item \textbf{Materializing implicit structures:} Querying and analyzing OCED benefits from explicitly materializing concepts and relations that otherwise are encoded implicitly in event or object attributes. For instance, object states (i.e., ``snapshots'', see Sect.~\ref{sec:challenges:objects_over_time})~\cite{broniewski2023,chu2022},  context knowledge such as connection and layout of objects describing components supporting a physical process in manufacturing or logistics~\cite{chu2022,cuprinsu2022,broniewski2023,DBLP:conf/bpm/SwevelsDF23}, or temporal and causal relations~\cite{esser_2020_multidimensional,DBLP:conf/er/KhayatbashiHJ23}. While these should not be stored in OCED for transport, it requires agreement on the implicit semantics of event, objects, and relations (see Sect.~\ref{sec:challenges:qualifiers}) for reliably constructing them.

    \item \textbf{Circular processes:} Use (or inference) of a consistent object identifier enables OCED to describe and track objects in circular processes where objects repeatedly return to the process over extended periods of time.~\cite{chu2022}

\end{enumerate}

%Links to tool/resources
\subsubsection*{Resources.}
OCED-PG is implemented as part of the open-source process mining library PromG (\url{https://github.com/promg-dev}) which provides functionality to develop process mining analyses over OCED in graph databases~\cite{DBLP:conf/icpm/SwevelsKF23_promg}. Schemas and transformation rules using OCED-PG are available for five public real-life datasets and one educational example~\cite{swevels_2023_8296559}. Further, a JSON-based OCED export from Konekti (Sect.~\ref{sec:implementation:konekti}) can be imported directly into OCED-PG as an EKG without any data transformation, see \url{https://github.com/PromG-dev/promg-konekti}.

%%% -------------------------------------------------------------
\subsection{Object-Centric Event Log (OCEL 2.0)}\label{sec:implementation:ocel2}
%%% -------------------------------------------------------------

OCEL, which stands for \textit{Object-Centric Event Log}, serves as the exchange format for Object-Centric Event Data (OCED) and is the foundation for a range of Object-Centric Process Mining (OCPM) techniques and over ten process mining tools and libraries \cite{Aalst_2023_ocpm_unraveling,ocel2_specification,DBLP:conf/icpm/KorenABA23}. 
OCEL 2.0 was released in 2023, extending OCEL 1.0 (released in 2020).
Appendix~\ref{app:ocel2} discusses the OCEL 2.0 meta-model (see Figure~\ref{fig:ocel2:full}).

\subsubsection*{OCEL 2.0 Storage Formats.}
There are different storage formats for OCEL 2.0: XML, JSON, and SQL. 
The detailed specifications for these formats can be found on the OCEL website: \url{https://www.ocel-standard.org/}.
Also, several event logs are provided in all three formats.
There are also publicly available tools to check the validity of an OCEL 2.0 dataset, 
e.g., an XML Schema, a JSON schema, and an SQL validator.

\subsubsection*{Lessons Learned.}
The most important lesson learned is to avoid adding concepts for which there are no analysis techniques 
or that allow for multiple ways of representing the same information. 
For example, there is already a trade-off between representing event-to-object relations (e.g.,  \texttt{observes}) and object-to-object relations. 
An order may have multiple items (object-to-object), and an event of type ``place order'' may include the order and its items (event-to-object). Further extending the meta-model will lead to even more trade-offs, e.g., the same information can be represented in many different, possibly redundant, ways.
We first need guidelines for using the existing concepts before adding new ones.

In the experiments and case studies with OCEL and dozens of process mining implementations using the Process Intelligence Graph (PIG) of Celonis (which uses a meta-model similar to OCEL 2.0 to store events and objects), we noted that
object-to-object relations are mostly used for filtering and querying, and event-to-object relations are mostly used for process discovery and conformance checking.
Due to the similarity, it is possible to load OCEL 2.0 into the Celonis ecosystem without any problems. 

These experiences suggest that, at this stage, there is no need for additional concepts.
XES also suffered from the problem that few of the extensions were actually being used. 
Therefore, it is much more important to support existing concepts well
and develop powerful process mining techniques (process discovery, conformance checking, predictive analytics) using both event-to-object and object-to-object relations. Concepts should only be added if there are mature analysis techniques exploiting these. 

\subsubsection*{Process Mining Tools and Libraries Supporting OCEL 2.0.}
\begin{itemize}
    \item OCEL 2.0 \textit{validators} using XML Schema, JSON schema, and an SQL checker, accessible via \url{https://www.ocel-standard.org/}.
    \item The web-based event log inspector for OCEL \textit{Ocelot}, accessible via \url{https://ocelot.pm/}.
    \item The \textit{OCPM} (Object-Centric Process Mining) tool, accessible via \url{https://www.ocpm.info/}. 
    \textit{OCPM} supports object-centric process discovery, object-centric conformance checking, and object-centric machine learning (using DFGs and Petri nets).
    \item The \textit{Process Mining for Javascript} (PM4JS) implementation (see \url{https://www.pm4js.org/}.    
    \item The \textit{OCELStandard} plugin for the \textit{ProM} framework (\url{https://promtools.org/}).
    \item The \textit{OCPA library} supporting object-centric process discovery, object-centric conformance checking, object-centric process enhancement, and object-centric process monitoring, accessible via \url{https://ocpa.readthedocs.io/}. 
    \item The \textit{Object-Centric Process Insights} ({OC$\pi$}) tool using OCPA. {OC$\pi$} supports object-centric process discovery and filtering, and provides elaborate support for object-centric variants. Download from \url{https://ocpi.ai/}.
    \item Connectors for Celonis, SAP, and Oracle (see \url{https://www.ocel-standard.org/}).
    \item The \textit{Object-Centric Process Querying} (OCPQ) tool to query OCEL 2.0 datasets (supporting the JSON, XML, and SQL formats), see \url{https://ocpq.aarkue.eu}.
    \item \textit{Process Mining for Python} (PM4Py) fully supports OCEL 2.0, including object-centric process discovery and object-centric conformance checking (using Petri nets, DFGs, and object graphs), accessible via \url{https://processintelligence.solutions/pm4py} and \url{https://github.com/pm4py/pm4py-core}.
\end{itemize}
The list of tools and libraries illustrates the interoperability achieved by adopting OCEL 2.0.

%%% -------------------------------------------------------------
\subsection{Stack't}\label{sec:implementation:stack-t}
%%% -------------------------------------------------------------
Stack't is an open-source data transformation tool designed to support data preparation for object-centric process mining in a dynamic context, i.e., allowing for continuously adding new data while accommodating changes in the process data architecture such as new object types, event types, or attributes~\cite{bosmans2024}.
The tool's development is driven by practical data-engineering considerations, drawing from industry experience of continuously extracting data from source systems, such as ERP and MES databases, for various analytical purposes.

\subsubsection*{Features.} Similar to Konekti, described in Section~\ref{sec:implementation:konekti}, we opt for a staggered approach by first mapping the process data to an intermediate data store (hub) and providing extractors to generate object-centric event logs in various formats from this. Below is an overview of the current core capabilities of the Stack't tool.
\begin{itemize}
    \item \textbf{Continuous Ingestion:} Supports append-only incremental batch processing of process data.
    \item \textbf{Interactive Visuals:} Provides interactive visualizations for exploratory data exploration
    \item \textbf{Export Capabilities:} Allows exporting to OCEL 2.0, DOCEL, and Neo4j graph database formats.
    \item \textbf{Import from OCEL 2.0:} Includes functionality to import event logs formatted according to OCEL 2.0.
\end{itemize}

Stack't is a modular and flexible data stack (in our implementation using DuckDB and dbt) in a Docker container, and can be integrated into existing data architectures. More information, together with the source code, can be found in the GitHub repository \url{https://github.com/LienBosmans/stack-t}.
%Stack't is not distributed as a stand-alone product, but rather as a data stack (DuckDB + dbt) in a box (Docker container). Its modular, open-source design allows seamless integration into existing data architectures. For instance, the example stack uses DuckDB as its database, but it can be replaced with any other analytical database or data lake with a query endpoint (data lake house). The data transformations, currently defined in dbt, can be transferred to the data pipeline tool already in use. This can increase data quality and save costs, by enabling the re-use of existing data pipelines between data sources and an analytical data platform, and reduces the number of applications and connections that need to be maintained by the data engineering team. As a trade-off, the barrier of entry for inexperienced data teams or business users is higher compared to an off-the-shelf tool with a user-friendly interface.

\subsubsection*{OCED-MM Changes.} As described in~\cite{bosmans2024}, Stack't adopts a flexible meta-model to extend the longevity of the data store. The implementation supports the OCED-MM Core Model, but also the OCED-MM Full Model and OCEL 2.0.
The flexibility manifests itself in two changes: object-to-object relationships are allowed to change over time by attaching a timestamp to the relationship qualifier, and direct relationships between events and object attribute value updates are used to store any known causal relationships between them and thus support many-to-many relationships. Storing process data that strictly adheres to the core model can be achieved by imposing restrictions, which are defined as additional data quality tests in the code:
\begin{enumerate}
    \item The table storing event-to-object-attribute-value relationships must be empty.
    \item The timestamp columns for object attribute values and object-to-object relationships must only contain NULL values.
    \item Relationships and attribute values must be uniquely identifiable using their foreign keys. The existing primary key column is kept, but must be supplemented with an additional check to ensure this is the case.
\end{enumerate}

%While it is relatively easy to come up with rules to map process data to a more strict format, the reverse is not true. By avoiding losing process information too soon during data preparation, no significant efforts are expected when future analysis techniques support additional concepts (e.g., dynamic object-to-object relations).
% The main changes to the OCED-MM (and mores specifically the OCEL 2.0 MM) are:
% \begin{enumerate}
%     \item Object-to-object relationships are allowed to change over time by attaching a timestamp to the relationship qualifier.
%     \item Direct relationships between events and object attribute value updates are used to store any known causal relationships between them and thus support many-to-many relationships.
% \end{enumerate}

\subsubsection*{Lessons Learned.}
\begin{enumerate}
    \item \textbf{Lack of publicly available real-life source data:} The absence of publicly available source data hinders the ability to validate the correctness of a source-to-target mapping implementation and assess the effect of several decisions regarding the conversion of raw data into an event log on aspects such as performance. %While there are some object-centric event logs available, a number of those are reconstructed from originally non-object-centric (flat) event logs. This could lead to having a skewed representation of reality if a majority of research is based on these.
    \item \textbf{Scalability and Maintainability in Dynamic Environments:} Designing data storage solutions for processes in dynamic or agile environments requires extra care. For example: even minor modifications such as renaming a type or attribute, can cascade into code-level impacts if not anticipated. More of these considerations, and how they are tackled by the proposed relational schema can be found in~\cite{bosmans2024}.
    \item \textbf{Dynamic Object-to-Object Relations:} The OCED-MM Core Model does not allow object-to-object relationships to evolve over time. Although visualizing such changes is complex, it remains essential for certain use cases. Consider an organizational chart for example: roles within a team might change. When, e.g., retroactively checking for irregularities, it would be important to know which relationship existed at certain points in time. A potential solution could be allowing dynamic relations between objects (see Sect.~\ref{sec:challenges:relations_as_objects}), but formulating guidelines on how additional restrictions such as static object-to-object relations should be handled when algorithms cannot handle dynamic ones as input. %This also applies to ambiguity that can result from having both timestamped object attribute values and timestamped events. We would prefer addressing this point based on the use-case, instead of never allowing to have both.
    \item \textbf{Standards and Definitions:} The lack of clear definitions for object-centric event (log) standards creates challenges, especially for those unfamiliar with the academic field. This was mainly apparent when trying to write connectors that can transform data from one format into another, as automating this requires the data to adhere to an unambiguous definition, and therefore knowing table and column names, as well as their data types, in advance (and for them to remain consistent over time). 
    \item \textbf{Data Anonymization:} We have tried to take into account the possible need for data anonymization with Stack't. While hiding the descriptions of types and attributes is rather straightforward, there is no current solution for masking sensitive information revealed through relationships.
\end{enumerate}

%%% -------------------------------------------------------------
\subsection{Further Known Implementations}\label{sec:implementation:further}
%%% -------------------------------------------------------------

Further implementations of OCED are under development.
\begin{itemize}
\item \emph{OCEDO} | a semantic-web-based ontology for core OCED defining the OCEDO namespace at \url{https://semsys.ai.wu.ac.at/ocedo/core} and algorithms for converting flat event logs into OCED via semantic technology \url{https://gitlab.isis.tuwien.ac.at/Ekaputra/ocedo}
\end{itemize}

%%% =============================================================
\section{Conclusion}
\label{sec:conclusion}
%%% =============================================================

The current consensus and state of discussion on identifying an object-centric event data format that can succeed XES can be summarized as follows.
\begin{enumerate}
    \item \textbf{Core concepts.} We identified the core concepts needed to represent object-centric event data that emerged from the intersection of multiple prior proposals and use cases. These core concepts are described in the OCED-MM Core Model in Sect.~\ref{sec:oced_core}.
    \item \textbf{Baseline interpretation and five implementations.} We provide a baseline interpretation for these core concepts in Sect.~\ref{sec:oced_core:interpretation} that is shared by all five existing OCED implementations, demonstrating basic viability of object-centric event data exchange.
\end{enumerate}
Subsequently, we (1) outline concrete steps that can be undertaken with the current OCED-MM Core Model and existing implementations, (2) summarize the challenges in realizing interoperability with OCED that need to be considered, and (3) outline general considerations towards standardizing OCED.

\subsubsection{Reliable basis for research and development.}
While the OCED-MM Core Model does not yet cover all practical requirements for data exchange in all use cases, and thus requires further clarification and (careful, limited) extension, it is adequate for the time being. Specifically, the model and the lessons learned documented in Sections~\ref{sec:challenges} and \ref{sec:implementations} provide a reliable basis for the process mining community to engage with OCED and related use cases, which include:
\begin{itemize}
\item Creating more object-centric event data sets to serve as realistic examples. This specifically includes creating or providing raw source data sets for conversion into OCED.
\item Making sure that existing and future implementations (see Sect.~\ref{sec:implementations}) can be connected, e.g., creating bridges and/or import/export functionality between OCEL 2.0 and OCED-PG and ensuring data imports and exports of different tools are compatible/compliant with each other.
\item Creating more OCED extractors for various source systems.
\item Identifying, documenting, and sharing process mining use cases that benefit
from the analysis of object-centric event data.
\item Researching, developing, and sharing process mining algorithms that consume
object-centric event data to address OCED-specific use cases.
\item Encouraging vendors to support OCED and object-centric process mining and educating students and users.
\end{itemize}
These results of these efforts will provide relevant information for a community-wide adoption of OCED and its standardization.

\subsubsection{Challenges in realizing interoperability with OCEDs.}
The diversity of lessons learned in the five independent implementations (see Sect.~\ref{sec:implementations}) reveals the limitations of the OCED-MM Core Model and the different design decisions made in extending or using OCED in particular ways. Current and further developments of OCED should be aware of the ambiguities in the OCED-MM Core Model and the implications of either design decision. 
\begin{enumerate}
    \item While the description of the OCED-MM Core Model allows for a reasonably unambiguous interpretation of each individual OCED concept, the description is \emph{not specific enough} for interpreting all \emph{combinations of concepts} unambiguously, e.g., how are multiple observations of an object represented (see Sect.~\ref{sec:challenges:objects_over_time}), standardizing \emph{qualifiers} to provide specific interpretation (see Sect.~\ref{sec:challenges:qualifiers}), and can object attributes be used to represent relations between objects (see Sect.~\ref{sec:challenges:relations}). This semantic ambiguity needs to be resolved to ensure that independent implementations of OCED can produce and consume object-centric event data with the same interpretation.
    
    \item The minimal concepts in OCED-MM Core Model do not support a number of use cases (see Sect.~\ref{sec:oced_core:limitations}). This requires either re-purposing existing concepts for additional use cases such as introducing artificial objects for relations, attributes, and other advanced concepts (see Sect.~\ref{sec:challenges:relations_as_objects}, \ref{sec:challenges:attributes_as_objects}, \ref{sec:challenges:other_as_objects}) or extending the OCED-MM with additional concepts (see Sect.~\ref{sec:challenges:strict_extensions}). Either form of extension introduces additional ambiguity in interpreting combinations of OCED concepts that need to be resolved (see Sect.~\ref{sec:challenges:cycles_ambiguity}), or deliberately be left out of scope of an OCED standard and deferred to later evolution of the standard or best practices as use cases mature.

    \item All mentioned ambiguities and limitations presented in this document can be resolved by taking design decisions wrt.\ representation and interpretation of the various OCED concepts. The existing OCED implementations presented in Sect.~\ref{sec:implementations} have done so, also including first steps in achieving interoperability (though with notable limitations).
\end{enumerate}

\subsubsection{Towards a community standard.}
As the current implementations of OCED differ in the scope and interpretation of OCED, there currently does not exist an \emph{eco-system} for producing and consuming OCED. 

The exhaustive exploration and discussion of OCED over the previous three years suggests that the overall space of OCED concepts, ambiguities, and design decisions for interpreting them is understood. This document can be understood as an inventory of all ambiguities and open design decisions for OCED. This suggests that the next step for realizing a community standard for OCED is to jointly review the known ambiguities design decisions and to form a community consensus of how to represent and interpret OCED concepts across a variety of use cases.

Forming consensus for representing and interpreting OCED around use cases thereby fundamentally requires considering the full life-cycle of object-centric event data. As each stage in the life-cycle has different constraints wrt. how data can be represented and processed, OCED may have to explicitly acknowledge different levels of consistency\footnote{\url{https://multiprocessmining.org/2022/10/26/data-storage-vs-data-semantics-for-object-centric-event-data/}}. For instance:
\begin{enumerate}
    \item \emph{Storage in source systems} (from which OCED is to be extracted) is optimized for usage, often lacking explicit, complete representations of events or objects, and lossy. Different types of events or objects may have fundamentally different storage representations.
    \item \emph{Extracting and providing OCED from source systems} requires conversion. Thereby some conversions may have prohibitive performance costs on the source system or lead to excessive data, e.g., translating an SAP ERP Change Table with 100 million records into events observing a modification of an object attribute.
    \item Other conversions may depend on a record's context, e.g., require to consider multiple records and additional domain knowledge, to provide a less ambiguous interpretation of an event or an object. This could be considered as a form of \emph{enrichment of OCED} that may depend on the \emph{analysis objectives}.
    \item Certain forms of resolving ambiguity in OCED interpretation may no longer be data (local) conversion tasks but \emph{genuine process mining tasks} providing use case-specific interpretations of the data, e.g., mining the Create-Read-Update-Delete life-cycle of a relation between objects or the identification of higher-level tasks from lower-level events.
    \item Finally, \emph{Process Mining solutions} working with OCED also have to realize OCED-compliant data structures and data stores that provide a standardized representation and interpretation of event data for all analysis and mining algorithms. These internal data stores may have higher requirements on how explicit various concepts and relations are represented and their consistency.
\end{enumerate}
The above non-exhaustive list illustrates a \emph{spectrum of different levels of ``strictness''} regarding representing and interpreting OCED that all originate from different constraints and requirements for the respective task and objective. 

Subsequent steps in standardizing OCED should explicitly consider these diverse levels of requirements and invite industry practitioners, vendors, and academic experts to jointly review design decisions wrt. supporting relevant use cases along the various stages of the OCED life-cycle.

The above considerations suggest that a community-supported OCED standard may have to explicitly support a (well-defined) spectrum of representations and interpretations of OCED along the data life-cycle. One possibility is to formulate well-defined \emph{levels} of OCED consistency and completeness. This could entail:
\begin{itemize}
\item Agreeing on the degree of inconsistency, i.e., kinds of ambiguity or incompleteness in representation and consistency, is allowed at a particular level (e.g., timestamped attribute values extracted from source system vs a minimum standard of unique representation of events and associated objects for process mining algorithms), and
\item identifying core principles for increasing the level of consistency and completeness of OCED by conversion, conventions, or extensions and the associated decisions in representation and interpretation of OCED.
\end{itemize}

\subsubsection*{Acknowledgments.} Thanks to the PADS PhDs and Postdocs that co-developed OCEL 2.0 and related tools, in particular 
Alessandro Berti,  Istvan Koren, Niklas Adams, Nina Graves,
Gyunam Park, Benedikt Knopp, Marco Pegoraro, Lukas Li{\ss},
Leah Tacke genannt Unterberg, Christopher Schwanen,  Aaron K{\"{u}}sters, Dina Kretzschmann, and Viki Peeva.\\
Thanks to the team from the Inco,FING,UdelaR (Uruguay) that participated in the implementation of OpenOCED related tools: Carolina Cortés, José Pedro De León, Maximiliano Jara and Camilo López.\\
Thanks to the Master students and PhD students at TU Eindhoven who contributed to OCED-PG and the case studies enabling it, in particular Stefan Esser, Ava Swevels, Eva Klijn, Maren Buermann, Vi Chu, Adam Broniewski, and Kadir Marangoz, as well as Francesca Zerbato for her feedback.\\
The contributions by TU Eindhoven are partially supported by AutoTwin EU GA n. 101092021.

\bibliographystyle{splncs04}
\bibliography{references}

\begin{thebibliography}{10}
\providecommand{\url}[1]{\texttt{#1}}
\providecommand{\urlprefix}{URL }
\providecommand{\doi}[1]{https://doi.org/#1}

\bibitem{DBLP:conf/sefm/Aalst19}
van~der Aalst, W.M.P.: Object-centric process mining: Dealing with divergence and convergence in event data. In: {\"{O}}lveczky, P.C., Sala{\"{u}}n, G. (eds.) Software Engineering and Formal Methods - 17th International Conference, {SEFM} 2019, Oslo, Norway, September 18-20, 2019, Proceedings. Lecture Notes in Computer Science, vol. 11724, pp. 3--25. Springer (2019). \doi{10.1007/978-3-030-30446-1\_1}, \url{https://doi.org/10.1007/978-3-030-30446-1\_1}

\bibitem{Aalst_2023_ocpm_unraveling}
van~der Aalst, W.M.P.: Object-centric process mining: Unraveling the fabric of real processes. Mathematics  \textbf{11}(12) (2023). \doi{10.3390/math11122691}, \url{https://www.mdpi.com/2227-7390/11/12/2691}

\bibitem{ocpn_fi_2020}
van~der Aalst, W.M.P., Berti, A.: {Discovering Object-Centric Petri Nets}. Fundamenta Informaticae  \textbf{175}(1-4),  1--40 (2020)

\bibitem{DBLP:journals/cim/AcamporaVSAGV17_xes}
Acampora, G., Vitiello, A., Stefano, B.N.D., van~der Aalst, W.M.P., G{\"{u}}nther, C.W., Verbeek, H.M.W.: {IEEE} 1849: The {XES} standard: The second {IEEE} standard sponsored by {IEEE} computational intelligence society [society briefs]. {IEEE} Comput. Intell. Mag.  \textbf{12}(2), ~4--8 (2017). \doi{10.1109/MCI.2017.2670420}, \url{https://doi.org/10.1109/MCI.2017.2670420}

\bibitem{ocel2_specification}
Berti, A., Koren, I., Adams, J.N., Park, G., Knopp, B., Graves, N., Rafiei, M., Li{\ss}, L., genannt Unterberg, L.T., Zhang, Y., Schwanen, C.T., Pegoraro, M., van~der Aalst, W.M.P.: {OCEL} (object-centric event log) 2.0 specification. CoRR  \textbf{abs/2403.01975} (2024). \doi{10.48550/ARXIV.2403.01975}, \url{https://doi.org/10.48550/arXiv.2403.01975}

\bibitem{bosmans2024}
Bosmans, L., Peeperkorn, J., Goossens, A., Lugaresi, G., Smedt, J.D., Weerdt, J.D.: Dynamic and scalable data preparation for object-centric process mining (2024), \url{https://arxiv.org/abs/2410.00596}

\bibitem{broniewski2023}
Broniewski, A.: Building a digital asset: An event knowledge graph approach for integrating data and persisting object-centric process mining analysis in baggage handling systems (2023)

\bibitem{DBLP:conf/bpm/Calegari023}
Calegari, D., Delgado, A.: A model-driven engineering perspective for the object-centric event data {(OCED)} metamodel. In: Weerdt, J.D., Pufahl, L. (eds.) Business Process Management Workshops - {BPM} 2023 International Workshops, Utrecht, The Netherlands, September 11-15, 2023, Revised Selected Papers. Lecture Notes in Business Information Processing, vol.~492, pp. 508--520. Springer (2023). \doi{10.1007/978-3-031-50974-2\_38}, \url{https://doi.org/10.1007/978-3-031-50974-2\_38}

\bibitem{chu2022}
Chu, V.: Using event knowledge graphs to model multi-dimensional dynamics in a baggage handling system (2022)

\bibitem{cuprinsu2022}
Cuprinsu, N.: Extending knowledge graphs to predict system and item behavior (2022)

\bibitem{esser_2020_multidimensional}
Esser, S., Fahland, D.: Multi-dimensional event data in graph databases. J. Data Semant.  \textbf{10},  109–141 (2021)

\bibitem{DBLP:reference/bdt/Fahland19}
Fahland, D.: Artifact-centric process mining. In: Sakr, S., Zomaya, A.Y. (eds.) Encyclopedia of Big Data Technologies. Springer (2019). \doi{10.1007/978-3-319-63962-8\_93-1}, \url{https://doi.org/10.1007/978-3-319-63962-8\_93-1}

\bibitem{DBLP:conf/er/KhayatbashiHJ23}
Khayatbashi, S., Hartig, O., Jalali, A.: Transforming event knowledge graph to object-centric event logs: {A} comparative study for multi-dimensional process analysis. In: Almeida, J.P.A., Borbinha, J., Guizzardi, G., Link, S., Zdravkovic, J. (eds.) Conceptual Modeling - 42nd International Conference, {ER} 2023, Lisbon, Portugal, November 6-9, 2023, Proceedings. Lecture Notes in Computer Science, vol. 14320, pp. 220--238. Springer (2023). \doi{10.1007/978-3-031-47262-6\_12}, \url{https://doi.org/10.1007/978-3-031-47262-6\_12}

\bibitem{DBLP:conf/icpm/KorenABA23}
Koren, I., Adams, J.N., Berti, A., van~der Aalst, W.M.P.: {OCEL} 2.0 resources - www.ocel-standard.org. In: van~der Werf, J.M.E.M., Cabanillas, C., Leotta, F., Genga, L. (eds.) Doctoral Consortium and Demo Track 2023 at the International Conference on Process Mining 2023 co-located with the 5th International Conference on Process Mining {(ICPM} 2023), Rome, Italy, October 27, 2023. {CEUR} Workshop Proceedings, vol.~3648. CEUR-WS.org (2023), \url{https://ceur-ws.org/Vol-3648/paper\_7195.pdf}

\bibitem{marangoz2023}
Marangoz, K.: Capturing multi-dimensional dynamics in a configuration management process through event knowledge graphs (2023)

\bibitem{DBLP:conf/bpm/SwevelsDF23}
Swevels, A., Dijkman, R.M., Fahland, D.: Inferring missing entity identifiers from context using event knowledge graphs. In: Francescomarino, C.D., Burattin, A., Janiesch, C., Sadiq, S. (eds.) Business Process Management - 21st International Conference, {BPM} 2023, Utrecht, The Netherlands, September 11-15, 2023, Proceedings. Lecture Notes in Computer Science, vol. 14159, pp. 180--197. Springer (2023). \doi{10.1007/978-3-031-41620-0\_11}, \url{https://doi.org/10.1007/978-3-031-41620-0\_11}

\bibitem{swevels_2023_8296559}
Swevels, A., Fahland, D.: Event data and semantic header for oced-pg (Aug 2023). \doi{10.5281/zenodo.8296559}, \url{https://doi.org/10.5281/zenodo.8296559}

\bibitem{DBLP:conf/icpm/SwevelsFM23_oced-pg}
Swevels, A., Fahland, D., Montali, M.: Implementing object-centric event data models in event knowledge graphs. In: Smedt, J.D., Soffer, P. (eds.) Process Mining Workshops - {ICPM} 2023 International Workshops, Rome, Italy, October 23-27, 2023, Revised Selected Papers. Lecture Notes in Business Information Processing, vol.~503, pp. 431--443. Springer (2023). \doi{10.1007/978-3-031-56107-8\_33}, \url{https://doi.org/10.1007/978-3-031-56107-8\_33}

\bibitem{DBLP:conf/icpm/SwevelsKF23_promg}
Swevels, A., Klijn, E.L., Fahland, D.: Object-centric process mining (and more) using a graph-based approach with promg. In: van~der Werf, J.M.E.M., Cabanillas, C., Leotta, F., Genga, L. (eds.) Doctoral Consortium and Demo Track 2023 at the International Conference on Process Mining 2023 co-located with the 5th International Conference on Process Mining {(ICPM} 2023), Rome, Italy, October 27, 2023. {CEUR} Workshop Proceedings, vol.~3648. CEUR-WS.org (2023), \url{https://ceur-ws.org/Vol-3648/paper\_9922.pdf}

\bibitem{DBLP:conf/caise/VerbeekBDA10_xes}
Verbeek, H.M.W., Buijs, J.C.A.M., van Dongen, B.F., van~der Aalst, W.M.P.: Xes, xesame, and prom 6. In: Soffer, P., Proper, E. (eds.) Information Systems Evolution - CAiSE Forum 2010, Hammamet, Tunisia, June 7-9, 2010, Selected Extended Papers. Lecture Notes in Business Information Processing, vol.~72, pp. 60--75. Springer (2010). \doi{10.1007/978-3-642-17722-4\_5}, \url{https://doi.org/10.1007/978-3-642-17722-4\_5}

\bibitem{DBLP:journals/cim/WynnAVS24_xes}
Wynn, M.T., van~der Aalst, W.M.P., Verbeek, H.M.W., Stefano, B.N.D.: The {IEEE} {XES} standard for process mining: Experiences, adoption, and revision [society briefs]. {IEEE} Comput. Intell. Mag.  \textbf{19}(1),  20--23 (2024). \doi{10.1109/MCI.2023.3333141}, \url{https://doi.org/10.1109/MCI.2023.3333141}

\bibitem{DBLP:conf/icpm/WynnLAACJV21_xes_survey}
Wynn, M.T., Lebherz, J., van~der Aalst, W.M.P., Accorsi, R., Ciccio, C.D., Jayarathna, L., Verbeek, H.M.W.: Rethinking the input for process mining: Insights from the {XES} survey and workshop. In: Munoz{-}Gama, J., Lu, X. (eds.) Process Mining Workshops - {ICPM} 2021 International Workshops, Eindhoven, The Netherlands, October 31 - November 4, 2021, Revised Selected Papers. Lecture Notes in Business Information Processing, vol.~433, pp. 3--16. Springer (2021). \doi{10.1007/978-3-030-98581-3\_1}, \url{https://doi.org/10.1007/978-3-030-98581-3\_1}

\end{thebibliography}

\appendix

%%% =============================================================
\section{Appendix: Original OCED-MM Base Model and OCED-MM Full Model}
\label{app:oced-mm-original}
%%% =============================================================

Before converging on the OCED-MM Core Model described in Sect.~\ref{sec:oced_core}, the OCED working group developed and published two predecessors: OCED-MM Based Model and OCED-MM Core Model. We include these here for reference for context of the discussion of limitations and extensions of the OCED-MM Core Model.

\subsection{Original OCED-MM Base Model}

The original OCED-MM Base Model developed by the OCED working group is shown in Fig.~\ref{fig:oced-mm:base}.

\begin{figure}
    \centering
    \includegraphics[width=\linewidth]{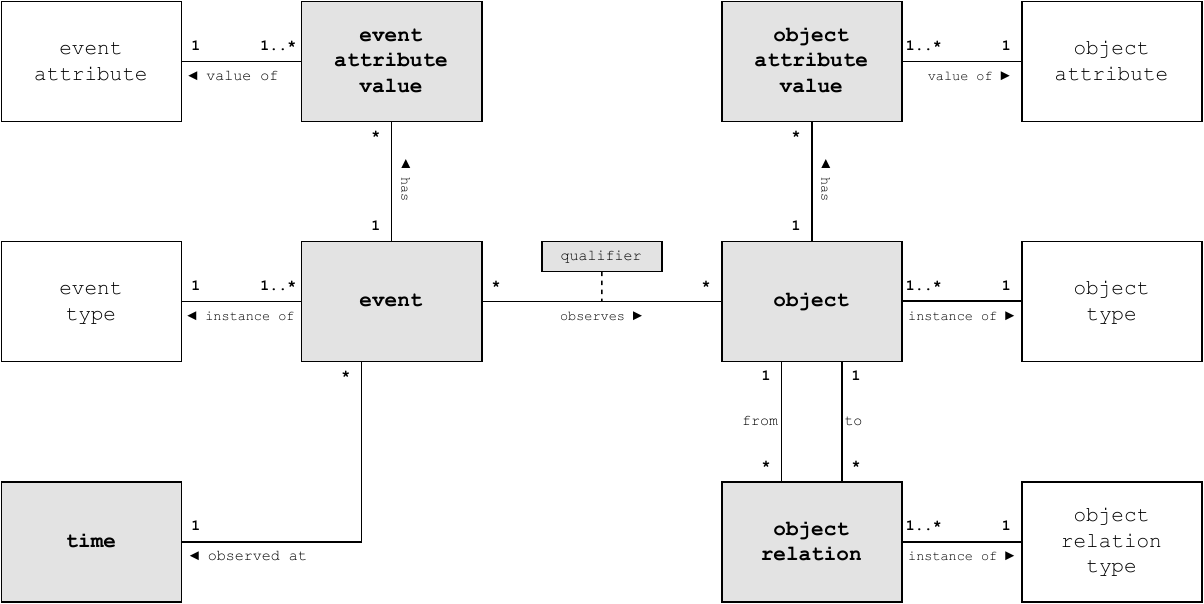}
    \caption{Original OCED-MM Base Model (published 22nd August 2022)}
    \label{fig:oced-mm:base}
\end{figure}

It differs from the OCED-MM Core Model described in Sect.~\ref{sec:oced_core} by modeling an \texttt{object relation} as a distinct identifiable entity of an \texttt{object relation type} and explicit \texttt{from} and \texttt{to} relationships referring to the objects that are related to each other. 

Section~\ref{sec:challenges:relations_as_objects} discusses the difference between these two choices of modeling object relations and how the OCED-MM Core Model of Sect.~\ref{sec:oced_core} can be extended to express object relations as distinct entities.

\subsection{Original OCED-MM Full Model}

The Full OCED-MM Model developed by the OCED working group is shown in Fig.~\ref{fig:oced-mm:base}.

\begin{figure}
    \centering
    \includegraphics[width=\linewidth]{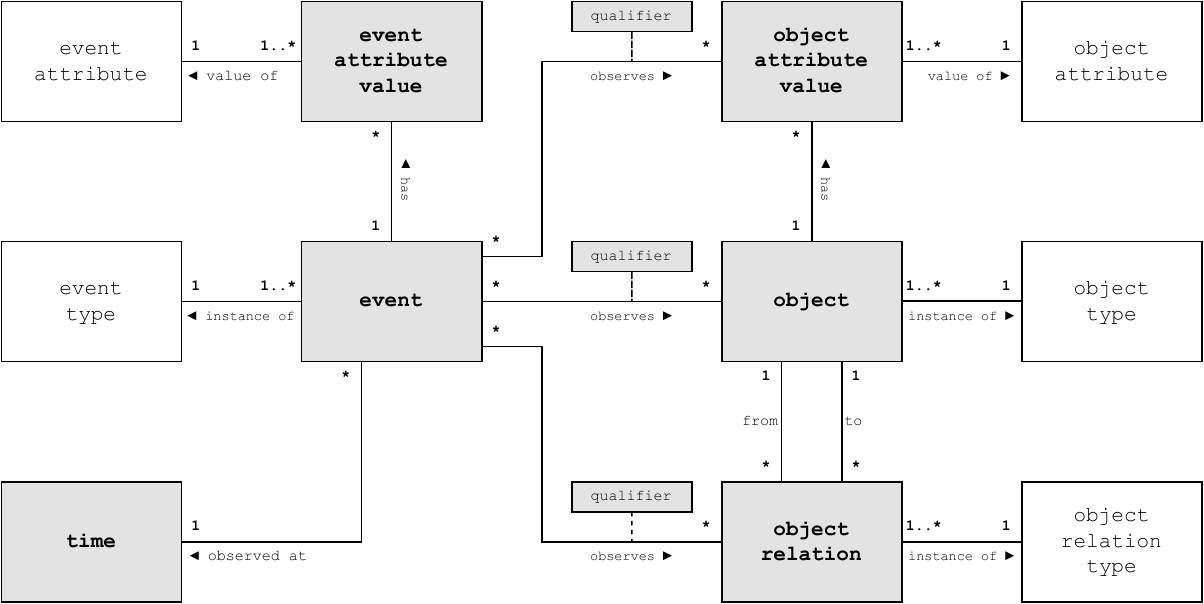}
    \caption{Original OCED-MM Base Model (published 22nd August 2022)}
    \label{fig:oced-mm:full}
\end{figure}

It further extends the OCED-MM Base Model by providing additional \emph{qualified} \texttt{observes} relationships from \emph{events} to, both, \texttt{object attribute values} and \texttt{object relations}.

\begin{enumerate}
    \item \textbf{event \texttt{observes} object attribute} | An \texttt{event} and an \texttt{object attribute value} can be related in a qualified (i.e., association class) manner, meaning their type of relationship is denoted. While a minimum set of qualifiers is predefined (\texttt{CREATE}, \texttt{MODIFY} and \texttt{DELETE}), additional qualifiers can be introduced as part of the data capture and used to reflect the semantics of the relationship. Each \texttt{object attribute value} can be involved in an arbitrary number of \texttt{events}, while each \texttt{event} can be related to an arbitrary number of \texttt{object attribute values}. This means, there can be events without object attribute values and vice-versa. In order to minimize the need to capture these relationships, any \texttt{object attribute value} that is not created explicitly (i.e., after it's object is in existence), is created implicitly with the \texttt{CREATE} of the \texttt{object} itself. When \texttt{objects} get deleted, all of their \texttt{object attribute values} are deleted implicitly.

    \item \textbf{event \texttt{observes} object relation} | An \texttt{event} and an \texttt{object relation} can be related in a qualified (i.e., association class) manner, meaning their type of relationship is denoted. While a minimum set of qualifiers is predefined (\texttt{CREATE}, \texttt{MODIFY}, and \texttt{DELETE}), additional qualifiers can be introduced as part of the data capture and used to reflect the semantics of the relationship. Each \texttt{object relation} can be involved in an arbitrary number of \texttt{events}, while each \texttt{event} can be related to an arbitrary number of \texttt{object relations}. This means, there can be \texttt{events} without \texttt{object relations} and vice-versa. When \texttt{objects} get deleted, all of their \texttt{object relations} are deleted implicitly.
\end{enumerate}

Section~\ref{sec:challenges:relations_as_objects} discusses how the OCED-MM Core Model of Sect.~\ref{sec:oced_core} can be extended to also express \texttt{observes} relations to object attributes and object relations.

This extension allows an event to refer to an object, an object attribute value, and/or an object relation. But an object can also refer to the same object attribute value and the same object relation. In case a log producer uses multiple of these linkage options simultaneously, cycles may be introduced leading to inconsistent data capture. Applying implicit semantics of the qualifiers of the \texttt{observes} relation and the \texttt{object relations}, e.g, deleting a parent object, allows to prevent such inconsistency. However, this is not enforced by the meta-model itself; see Sect.~\ref{sec:challenges:cycles_ambiguity}.

%%% =============================================================
\section{Appendix: OCEL 2.0}
\label{app:ocel2}
%%% =============================================================

OCEL 1.0 \textit{Object-Centric Event Log} was released in 2020, prior to the standardization process described in Section~\ref{sec:path_to_oced}. OCEL 1.0 was based on a number of attempts to standardize such event data in the period 2015-2019. See, for example, the eXtensible Object-Centric (XOC) event log format and a variety of artifact-centric formats (e.g., proclets).
Based on the limited adoption and support for these formats, OCEL 1.0 started deliberately simple, with a focus on object-centric discovery and conformance-checking techniques using only event-to-object (E2O) relationships \cite{Aalst_2023_ocpm_unraveling,ocpn_fi_2020}. OCEL 2.0, released in 2023, extends OCEL 1.0, leveraging experiences gathered while developing and applying these OCPM techniques \cite{Aalst_2023_ocpm_unraveling,ocel2_specification,DBLP:conf/icpm/KorenABA23}. 
In OCEL 2.0, Object-to-Object (O2O) relationships were added, next to qualifiers for both E2O and O2O relationships. Figure~\ref{fig:ocel2:full} shows the OCEL 2.0 meta-model.
The design process for the original OCED-MM Base Model and OCED-MM Full Model (see Sect.~\ref{sec:path_to_oced}) explored a number of design decisions corresponding to variations of the OCED 2.0 meta-model \cite{Aalst_2023_ocpm_unraveling,ocel2_specification,DBLP:conf/icpm/KorenABA23}.
\begin{figure}
    \centering
    \includegraphics[width=\linewidth]{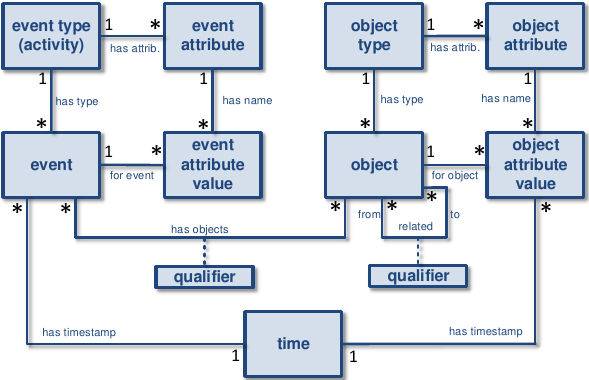}
    \caption{The full OCEL 2.0 meta-model \cite{Aalst_2023_ocpm_unraveling,ocel2_specification,DBLP:conf/icpm/KorenABA23} extending OCEL 1.0 and the OCED-MM Base Model}
    \label{fig:ocel2:full}
\end{figure}

Figure~\ref{fig:ocel2:full} shows the OCEL 2.0 meta-model which varies from and extends the original OCED-MM Base Model in various ways. 
The main differences are the ability to define the possible attributes per object and event type, and the ability to timestamp object attribute values. 
For an explanation of these concepts, we refer to Section \ref{sec:implementation:ocel2} and \cite{ocel2_specification,DBLP:conf/icpm/KorenABA23,Aalst_2023_ocpm_unraveling}.
\begin{figure}
    \centering
    \includegraphics[width=\linewidth]{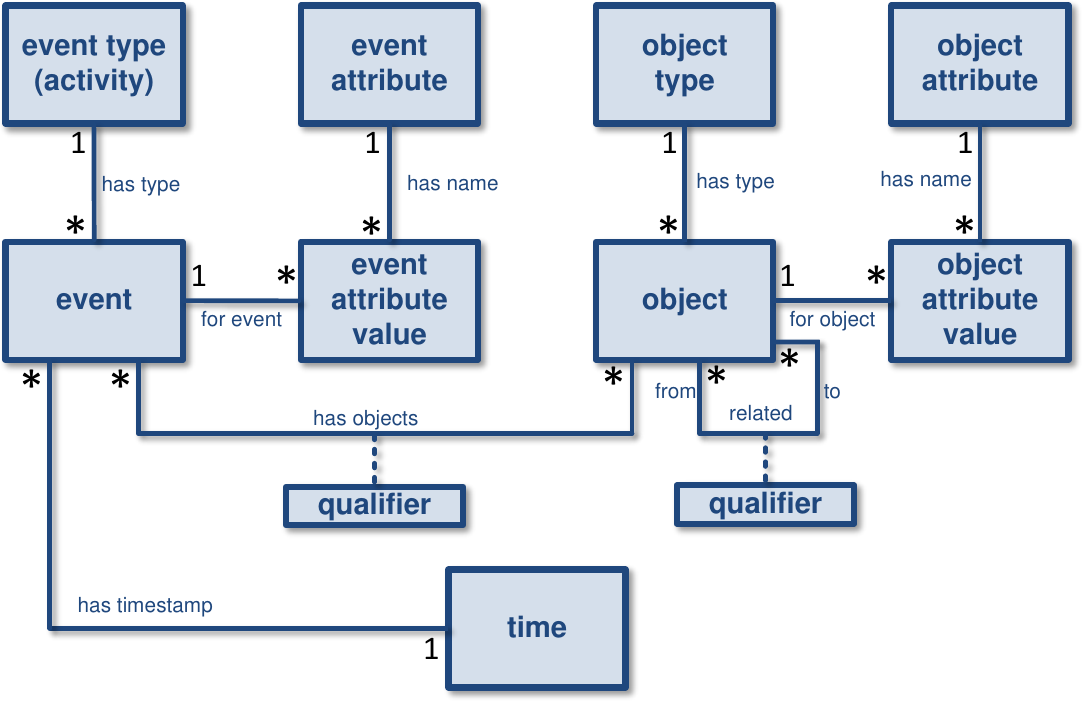}
    \caption{The reduced OCEL 2.0 meta-model without timed object attributes and without the ability to specify possible attributes per object and event type}
    \label{fig:ocel2:reduced}
\end{figure}

Omitting the \texttt{has attribute} relationships between attributes and types for events and objects, and the \texttt{has timestamp} relationships 
from \texttt{object} \texttt{attribute} \texttt{values} to \texttt{time} results in the reduced OCEL 2.0 meta-model shown in Fig.~\ref{fig:ocel2:reduced} which is similar to the OCED-MM Core Model of Sect.~\ref{sec:oced_core} (up to naming of relationships).
Note that omitting the \texttt{has attribute} relationships makes it more difficult to store the data in a relational database. When events and objects of the same type may have arbitrary attributes, it is impractical to store these in a relational database. See the OCEL 2.0 SQLite format \cite{ocel2_specification} for details.

%Sections~\ref{sec:challenges:strict_extensions} and \ref{sec:challenges:cycles_ambiguity} discuss, among others, the challenges of extending the OCED-MM Core Model with the concepts proposed in OCEL 2.0.

\end{document}